\documentclass[amsmath,amssymb,superscriptaddress,nobalancelastpage,prb,twocolumn,]{revtex4-1}

\AtBeginDocument{%
    \newwrite\bibnotes
    \def\bibnotesext{Notes.bib}
    \immediate\openout\bibnotes=\jobname\bibnotesext
    \immediate\write\bibnotes{@CONTROL{REVTEX41Control}}
    \immediate\write\bibnotes{@CONTROL{%
    apsrev41Control,author="08",editor="1",pages="1",title="0",year="1"}}
     \if@filesw
     \immediate\write\@auxout{\string\citation{apsrev41Control}}%
    \fi
}%

\usepackage{graphicx}
\usepackage{varioref}
\usepackage{xr-hyper}
\usepackage{xcolor}
\usepackage{nicefrac}
\usepackage{xfrac}
\usepackage{hyperref}
\hypersetup{colorlinks,linkcolor=blue,urlcolor=blue,citecolor=blue}
\usepackage{ulem}
\usepackage{siunitx}
\usepackage{graphicx}
\usepackage{dcolumn}
\usepackage{bm}
\usepackage{braket}
\usepackage{wasysym}
\usepackage{textcomp}

\newcommand{\lsco}{{La}$_{1.88}${Sr}$_{0.12}${CuO}$_4$}

\begin{document}

\title{Uniaxial Pressure Induced Stripe Order Rotation in \lsco}

\author{Qisi~Wang}
\email{qisiwang@physik.uzh.ch}
\affiliation{Physik-Institut, Universit\"{a}t Z\"{u}rich, Winterthurerstrasse 190, CH-8057 Z\"{u}rich, Switzerland}

\author{K.~von~Arx}
\affiliation{Physik-Institut, Universit\"{a}t Z\"{u}rich, Winterthurerstrasse 
190, CH-8057 Z\"{u}rich, Switzerland}
\affiliation{Department of Physics, Chalmers University of Technology, SE-412 96 G\"{o}teborg, Sweden}

\author{D.~G.~Mazzone}
\affiliation{Laboratory for Neutron Scattering and Imaging, Paul Scherrer Institut, CH-5232 Villigen PSI, Switzerland}

\author{S.~Mustafi}
\affiliation{Physik-Institut, Universit\"{a}t Z\"{u}rich, Winterthurerstrasse 
190, CH-8057 Z\"{u}rich, Switzerland}

\author{M.~Horio}
\affiliation{Physik-Institut, Universit\"{a}t Z\"{u}rich, Winterthurerstrasse 
190, CH-8057 Z\"{u}rich, Switzerland}

\author{J. K\"{u}spert}
\affiliation{Physik-Institut, Universit\"{a}t Z\"{u}rich, Winterthurerstrasse 190, CH-8057 Z\"{u}rich, Switzerland}

\author{J.~Choi}
\affiliation{Physik-Institut, Universit\"{a}t Z\"{u}rich, Winterthurerstrasse 
190, CH-8057 Z\"{u}rich, Switzerland}

\author{D.~Bucher}
\affiliation{Physik-Institut, Universit\"{a}t Z\"{u}rich, Winterthurerstrasse 
190, CH-8057 Z\"{u}rich, Switzerland}

\author{H.~Wo}
\affiliation{State Key Laboratory of Surface Physics and Department of Physics, Fudan University, Shanghai 200433, China}

\author{J.~Zhao}
\affiliation{State Key Laboratory of Surface Physics and Department of Physics, Fudan University, Shanghai 200433, China}

\author{W.~Zhang}
\affiliation{Photon Science Division, Swiss Light Source, Paul Scherrer Institut, CH-5232 Villigen PSI, Switzerland}

\author{T.~C.~Asmara}
\affiliation{Photon Science Division, Swiss Light Source, Paul Scherrer Institut, CH-5232 Villigen PSI, Switzerland}

\author{Y.~Sassa}
\affiliation{Department of Physics, Chalmers University of Technology, SE-412 96 G\"{o}teborg, Sweden}

\author{M.~M\aa nsson}
\affiliation{Department of Applied Physics, KTH Royal Institute of Technology, SE-106 91 Stockholm, Sweden}

\author{N.~B.~Christensen}
\affiliation{Department of Physics, Technical University of Denmark, DK-2800 Kongens Lyngby, Denmark}

\author{M.~Janoschek}
\affiliation{Physik-Institut, Universit\"{a}t Z\"{u}rich, Winterthurerstrasse 190, CH-8057 Z\"{u}rich, Switzerland}
\affiliation{Laboratory for Neutron and Muon Instrumentation, Paul Scherrer Institut, CH-5232 Villigen PSI, Switzerland}

\author{T.~Kurosawa}
\affiliation{Department of Physics, Hokkaido University - Sapporo 060-0810, 
Japan}
 
\author{N.~Momono}
\affiliation{Department of Physics, Hokkaido University - Sapporo 060-0810, 
Japan}
\affiliation{Department of Applied Sciences, Muroran Institute of Technology, Muroran 050-8585, Japan}

\author{M.~Oda}
\affiliation{Department of Physics, Hokkaido University - Sapporo 060-0810, 
Japan}

\author{M.~H.~Fischer}
\affiliation{Physik-Institut, Universit\"{a}t Z\"{u}rich, Winterthurerstrasse 190, CH-8057 Z\"{u}rich, Switzerland}

\author{T.~Schmitt}
\affiliation{Photon Science Division, Swiss Light Source, Paul Scherrer Institut, CH-5232 Villigen PSI, Switzerland}

\author{J.~Chang}
\email{johan.chang@physik.uzh.ch}
\affiliation{Physik-Institut, Universit\"{a}t Z\"{u}rich, Winterthurerstrasse 190, CH-8057 Z\"{u}rich, Switzerland}

\maketitle

\begin{figure*}[t]
\center{\includegraphics[width=0.95\textwidth]{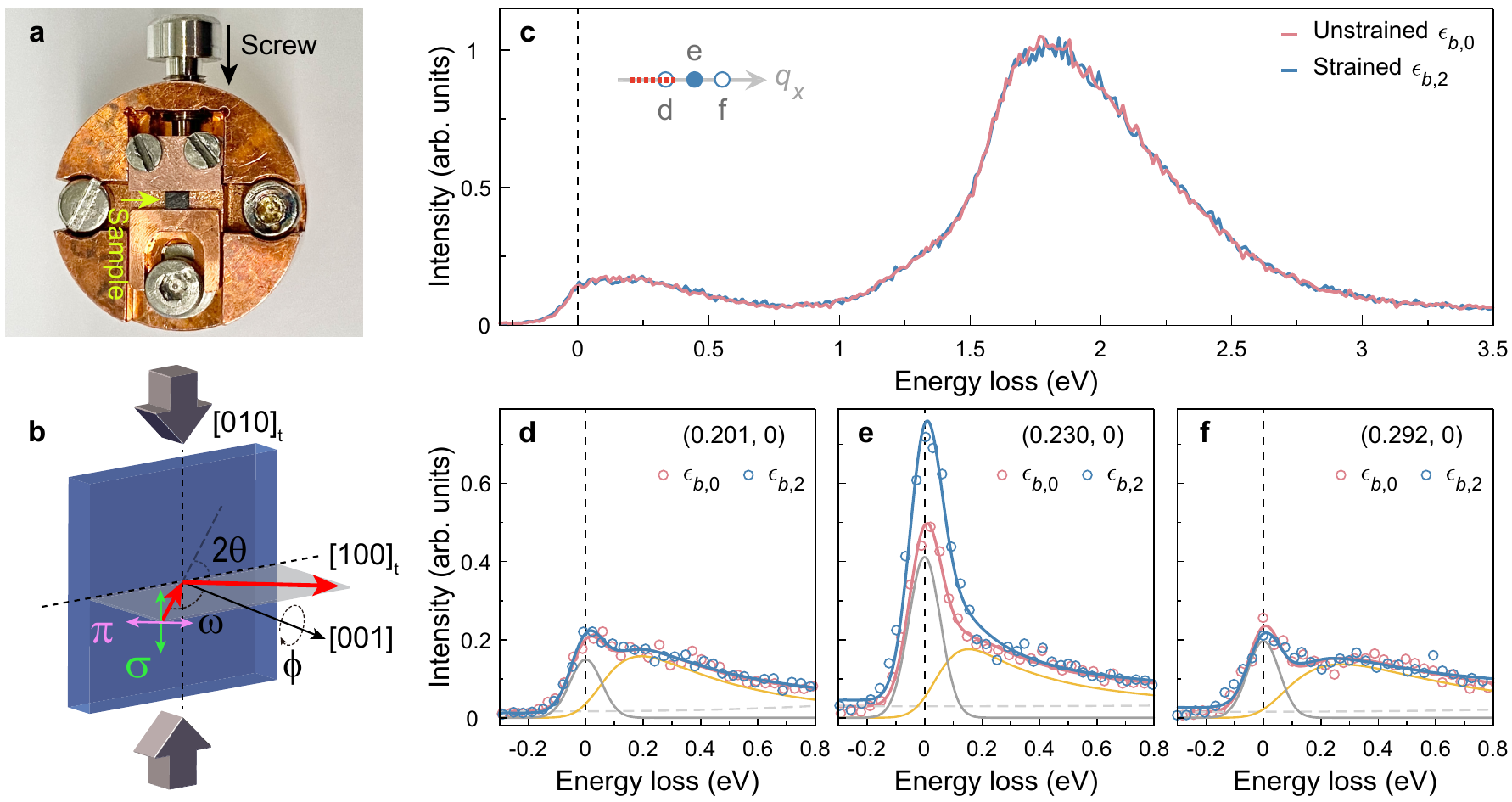}} 	
\caption{\textbf{Uniaxial strain tuning of charge-stripe order in \lsco.} \textbf{a} Photograph of the strain device---sample (black rectangle) and \textit{in-situ} mechanical screw mechanism are indicated with arrows. \textbf{b} Illustration of the scattering geometry with respect to the uniaxial strain application. Vertical ($\sigma$) and horizontal ($\pi$) linear light polarisations are indicated  with respect to the scattering plane.
\textbf{c} RIXS spectra including high-energy excitations
recorded with (blue, $\epsilon_{b,2}$) and without (red, $\epsilon_{b,0}$) strain for $T\approx28$~K. The inset indicates the momentum along $\mathbf{Q}=(q_\parallel,0)$.
Red dotted line reveals the momentum integration used for RIXS spectra in \textbf{c} and circles indicate the momenta for the RIXS spectra in \textbf{d-f}.
\textbf{d-f} Low-energy part of RIXS spectra recorded with and without strain for momenta as indicated. Line profiles in \textbf{d-f} represent fits including a polynomial background (grey dashed line), a damped harmonic oscillator to model the paramagnetic contribution (orange line), and a Gaussian line shape covering the elastic scattering (grey solid line). The intensity is given in arbitrary units.}
\label{fig1}
\end{figure*}

\begin{figure*}[t]
\center{\includegraphics[width=0.95\textwidth]{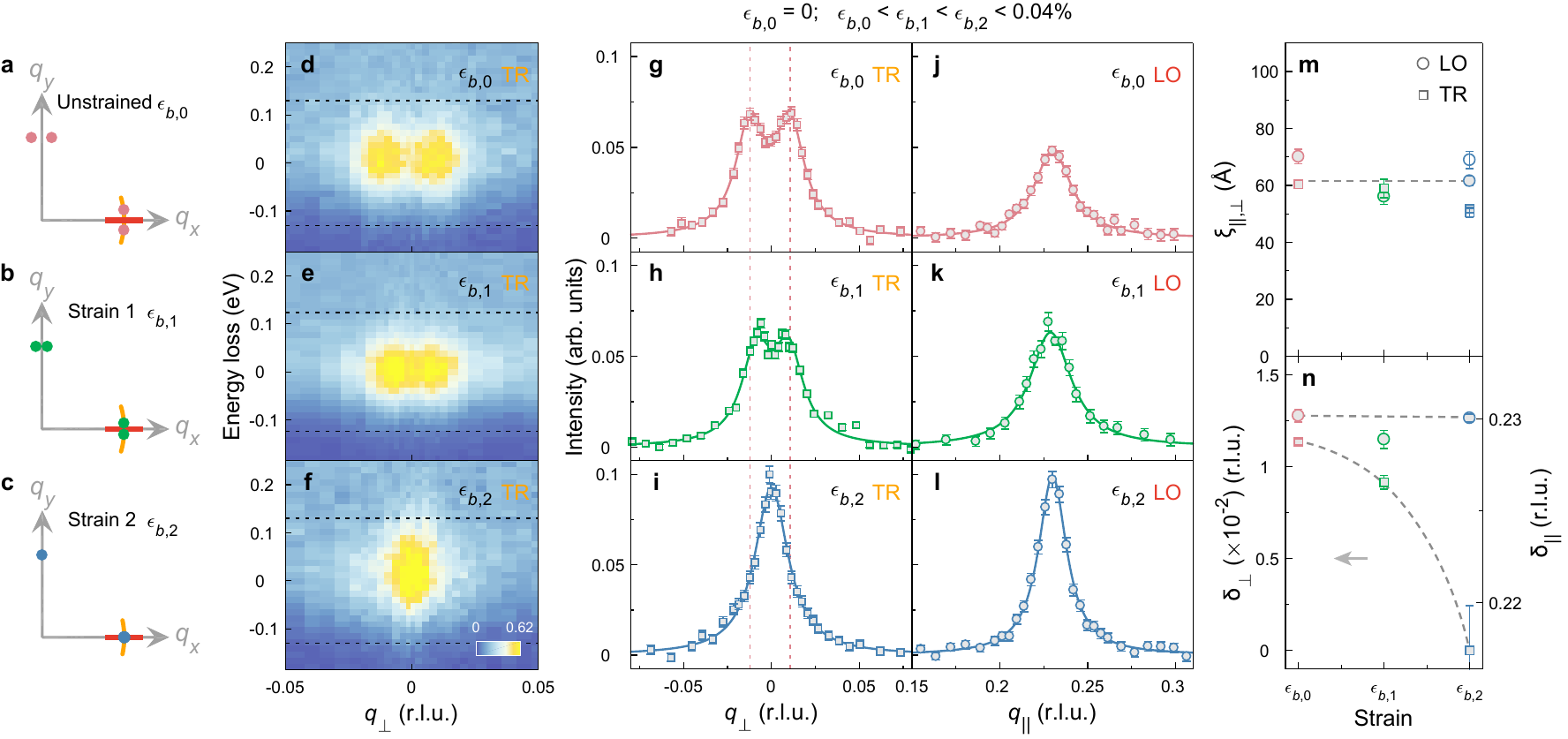}} 	
\caption{\textbf{Evolution of charge-stripe order structure in \lsco~under uniaxial strain.}
\textbf{a-c} Illustration of the stripe order diffraction pattern as uniaxial pressure is increased. The splitting of stripe order peaks in \textbf{a} and \textbf{b} is exaggerated for clarity.
Red and orange thick lines denote longitudinal (LO) and rocking (approximately transverse, TR)  scans, respectively shown in \textbf{g-l} for $T=28$~K~$\approx T_c$. 
\textbf{d-f} RIXS intensity maps as a function of energy and $q_\perp$, from which scans in \textbf{g-i} are obtained. Black dashed lines, in \textbf{d-f}, mark the energy integration window of elastic intensity. Red dashed lines, in \textbf{g-i}, indicate the fitted peak positions of the transverse scan at zero strain.
\textbf{m} Correlation length and \textbf{n} incommensurability along longitudinal and transverse directions extracted from fits (solid lines in \textbf{g-l}) as described in the main text and Supplementary Information. Open symbols in \textbf{m} denote results obtained on a repeated measurement with strain value comparable to strain 2. Error bars in \textbf{g-l} and \textbf{m, n} are set by counting statistics and standard deviation of fittings, respectively.
Uniaxial pressure is increased from strain 1 ($\epsilon_{b,1}$) to strain 2 ($\epsilon_{b,2}$) by mechanically turning the screw pressing onto the sample. Strain 2 corresponds to the strain value for data shown in Fig.~\ref{fig1}.}
\label{fig2}
\end{figure*} 

\begin{figure*}[t]
\center{\includegraphics[width=0.95\textwidth]{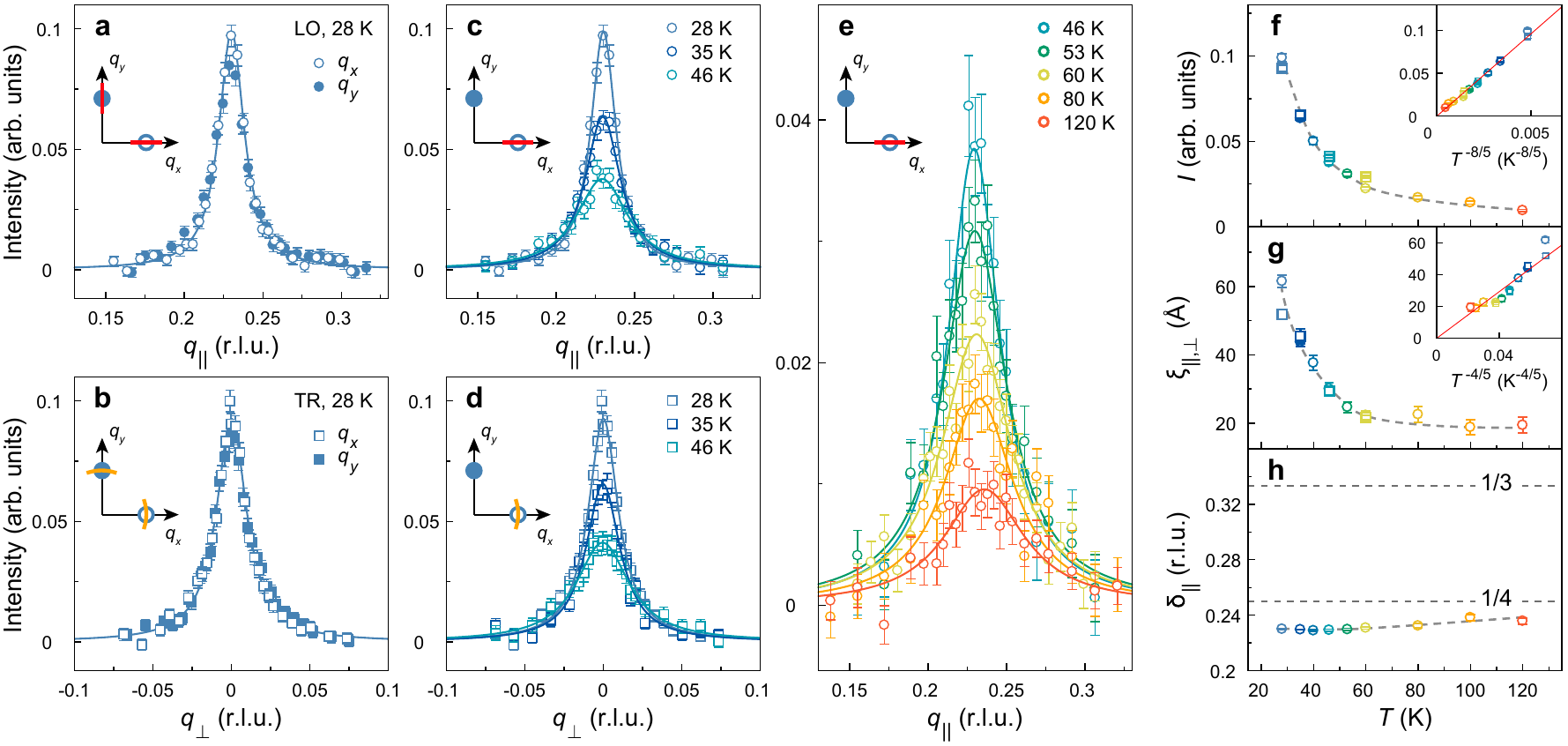}} 
\caption{\textbf{Temperature evolution of lattice pinned charge-stripe order in \lsco.} 
\textbf{a} Longitudinal (circles) and \textbf{b} transverse (square) scans through the stripe ordering vectors at $T\approx T_c$. \textbf{c-e} Temperature dependence of the stripe order diffraction peaks. Solid lines in \textbf{a-e} are Lorentzian fits from which amplitude, correlation length and incommensurability are inferred and plotted in \textbf{f-h} versus temperature. Circles and squares in \textbf{f-h} denote fitting results from the longitudinal and transverse scans, respectively. Error bars in \textbf{a-e} and \textbf{f-h} are set by counting statistics and standard deviation of fittings, respectively. Dashed lines in \textbf{f-g} are guides to the eyes. The two horizontal dashed lines in \textbf{h} indicate lattice commensurate values. Red solid lines in the insets of \textbf{f} and \textbf{g} are linear fits.}
\label{fig3}
\end{figure*}

\noindent\textbf{Static stripe order is detrimental to superconductivity. Yet, it has been proposed that transverse stripe fluctuations may enhance the inter-stripe Josephson coupling and thus promote superconductivity. Direct experimental studies of stripe dynamics, however, remain difficult. From a strong-coupling perspective, transverse stripe fluctuations are realized in the form of dynamic ``kinks"---sideways shifting stripe  sections. 
Here, we show how modest uniaxial pressure tuning reorganizes directional kink alignment. Our starting point is {La}$_{1.88}${Sr}$_{0.12}${CuO}$_4$, where transverse kink ordering results in a rotation of stripe order away from the crystal axis. Application of mild uniaxial pressure changes the ordering pattern and pins the stripe order to the crystal axis. This reordering occurs at a much weaker pressure than that to detwin the stripe domains and suggests a rather weak transverse stripe stiffness. Weak spatial stiffness and transverse quantum fluctuations are likely key prerequisites for stripes to coexist with superconductivity.} 
\vspace{1mm}

\noindent In the cuprates, stripes of doped holes---forming string-like antiferromagnetic domain walls that periodically modulate the charge density---have been both theoretically proposed~\cite{MachidaPhysC1989,ZaanenPRB1989,EmeryPhysC1993,CastellaniPRL1995,HellbergPRL1997,SeiboldPRB1998,HuangSci2017,ZhengSci2017} and experimentally revealed~\cite{TranquadaNat1995}.
The interplay between stripes and superconductivity is crucial~\cite{EmeryPhysC1994}.
Whereas static stripe order appears detrimental~\cite{ChangPRB2008,HuckerPRB2011}, fluctuating stripes may be favourable for superconductivity~\cite{KivelsonNat1998,KivelsonRMP2003,ScalapinoRMP2012}. 
Transverse stripe fluctuations, for example, have been theoretically suggested to promote superconductivity by enhancing the Josephson coupling between stripes~\cite{KivelsonNat1998}.
While such meandering motions of stripes are driven by zero-point fluctuations~\cite{KivelsonNat1998}, they are also subject to a finite spacial stiffness, stemming from Coulomb repulsion and the underlying crystal lattice that defines the direction of the stripe order. 
In systems like the cuprates~\cite{TranquadaNat1995}, nickelates~\cite{TranquadaPRL1994} and manganites~\cite{ShimomuraPRL1999,Vasiliu-DolocPRL1999}, charge-stripe order is thus usually pinned either parallel or diagonally to a principal atomic lattice axis.

In a strong-coupling picture, meandering stems from transverse excitations in the form of kinks shifting the stripes by an integer number of atomic units~\cite{BoschPRB2001,ZaanenPRB1998}. On a macroscopic level, such kinks depin the stripe from the lattice. It has been suggested that the transverse stripe fluctuations have a crucial effect on the competition between charge order and superconductivity, and lead to a rich phase diagram featuring an electronic solid, an isotropic phase, and in between liquid crystal states coexisting with superconductivity~\cite{KivelsonNat1998}. However, direct experimental studies of charge stripe dynamics remain challenging~\cite{MitranoSciAdv2019}. As yet, little is known about the lattice pinning potential and the transition from static to fluctuating meandering stripes.

Among hole-doped cuprate compounds, \lsco~(LSCO) is unique because the orthorhombic lattice distortion is diagonal to the stripes, providing a less compatible lattice ``host".
In LSCO, charge order is manifested by eight satellite reflections at $\mathbf{Q}=\bm{\mathrm{\tau}}+\mathbf{q_i}$~\cite{CroftPRB2014,Christensen14}. Here, $\bm{\mathrm{\tau}}$ is a fundamental Bragg peak and $\mathbf{q_{1,2}}=\pm(\delta_\parallel,\delta_\perp)$ with $\delta_\parallel\approx1/4$. 
The transverse incommensurability $\delta_\perp\approx0.011$ is far beyond the expectation from orthorhombic twinning~\cite{ThampyPRB2014,KimuraPRB2000,WakimotoJPSJ2006}.
The remaining six reflections appear at mirror ($q_y\rightarrow -q_y$) and rotation ($q_x\rightarrow q_y$) symmetric equivalent positions. 
In the strong-coupling picture, charge-stripe order is locally commensurate but with the possibility of phase jumping~\cite{McMillanPRB1976,BoschPRB2001,ZaanenPRB1998,EskesPRB1998}.
The modulation $\delta_\perp$ is understood via kink ordering that effectively rotates the stripes away from the principle crystal axes. LSCO is thus a unique example of charge order ``unlocked" from the lattice. As such, LSCO can be viewed as an intermediate stepping stone between statically  pinned  and fluctuating stripes.

Here, we study the transverse pinning properties of the charge-stripe order. For this purpose, we performed a resonant inelastic x-ray scattering (RIXS) experiment employing uniaxial pressure application. Our setup enables weak \textit{in-situ} compressive strain along a copper-oxygen bond direction. We show how modest strain application gradually pins the stripe order to the crystal axis along the copper-oxygen bond direction, and thus demonstrate that uniaxial pressure allows tuning of kink ordering in LSCO. We find that the lattice pinning potential is weaker than that to detwin the stripe order. This suggests that, at least in LSCO, transverse stripe fluctuations possess an energy scale relevant for the ground state properties. The recent demonstration of uniaxial pressure tuning of electronic instabilities in combination with RIXS~\cite{KimPRL2021,Lu21} opens a new avenue for spectroscopy studies of quantum materials. \\

\noindent\textbf{Results} \\
\noindent\textbf{Uniaxial pressure effect.}
Fig.~\ref{fig1} shows RIXS spectra recorded along the $(q_\parallel,0)$ direction with and without strain application. The spectra reveal an elastic and low-energy ($\lesssim 1$~eV) paramagnetic contribution and strong $dd$ excitations at higher energy. The $dd$ and the spin excitations show no discernible strain effect. This is in contrast to strain experiments on films.
In La$_2$CuO$_4$ thin films, strain $\epsilon=(a-a_0)/a_0$ of the order ${\sim}1\%$ yields a pronounced change of the $dd$ excitations. A modification of the $dd$ profile is easily detectable even with strain of $\epsilon\sim 0.1\%$ (ref.~\cite{IvashkoNC2019}). Similar results are reported on LaCoO$_3$ films~\cite{WangPRB2019}. Furthermore, the electronic and magnetic excitations in Sr$_2$IrO$_4$ thin films are sensitive to strain in the order of ${\sim}0.2\%$~\cite{ParisPNAS2020}.
Absence of uniaxial pressure effects in our experiment
suggests that the crystal field environment is only marginally modified~\cite{SalaNJP2011}. 
This is consistent with our strain calibration using x-ray diffraction (XRD) that yields an upper bound of the $c$-axis lattice expansion $\epsilon_c=(c-c_0)/c_0\lesssim 0.015\%$ (see Methods, Supplementary Note~3 and Supplementary Fig.~2), which is at least an order of magnitude weaker than the maximum strain applied in ref.~\cite{Choi20}. Applying Poisson's ratio~\cite{MeyerAPL2015,Choi20}, we estimate the upper bound of the in-plane compressive strain $-\epsilon_b=(b-b_0)/b_0$ applied to be $\epsilon_b\lesssim 0.04\%$. Data in our work were obtained under three strain conditions with $\epsilon_{b,0}=0$ and $0<\epsilon_{b,1}<\epsilon_{b,2}<0.04\%$.

This strain limit stems from the \textit{in-situ} operational screwdriver that provides finite mechanical force due to its magnetic coupling mechanism.
Our strain cell (Fig.~\ref{fig1}a) thus generates very modest uniaxial pressure. Away from $\mathbf{Q}=(\delta_\parallel,\pm\delta_\perp)$, strain has virtually no effect on the elastic and spin excitation scattering channels. The spectra obtained with and without uniaxial pressure are indistinguishable (Fig.~\ref{fig1}d,f). At $\mathbf{Q}=(\delta_\parallel,0)$ by contrast, elastic scattering is significantly increased after application of pressure (Fig.~\ref{fig1}e).\\

\noindent\textbf{Stripe order (de)pinning.}
RIXS spectra are fitted by modeling elastic and magnon scattering with a Gaussian and a damped harmonic oscillator functional form, respectively~\cite{WangPRL2020} (Fig.~\ref{fig1}d-f). The Gaussian width is fixed to the instrumental resolution.
Fig.~\ref{fig2}(d-f) displays RIXS intensity as a function of momentum and energy loss. Elastic intensity is obtained by integrating the spectral weight around zero energy within $\pm$FWHM energy window (black dashed lines, see Methods for details).
The resulting longitudinal $(q_\parallel,0)$ and transverse $(\delta_\parallel,q_\perp)$ scans are plotted in Fig.~\ref{fig2}(g-l).
Under ambient pressure condition, elastic scattering peaks appear at $(\delta_\parallel,\pm\delta_\perp)$~\cite{ThampyPRB2014,IvashkoPRB2017,WenNC2019}, resulting in a double peak structure in the transverse scan (Fig.~\ref{fig2}d,g). 
Upon application of uniaxial pressure, $\delta_\parallel$ remains unchanged (Fig.~\ref{fig2}(j-l)). By contrast, the transverse incommensurability $\delta_\perp$ is highly sensitive to uniaxial pressure and quickly vanishes upon pressure application (Fig.~\ref{fig2}(g-i)). This results in a transverse scan that features a single peak structure centred around $(\delta_\parallel,0)$.
The same effect is found along the perpendicular copper-oxygen bond direction (Fig.~\ref{fig3}a,b). Modest uniaxial pressure thus generates a twinned charge order structure with ordering vectors $(\delta_\parallel,0)$ and $(0,\delta_\parallel)$. As shown in Supplementary Fig.~4, the stripe order remains pinned after releasing uniaxial pressure. 
It is possible that uniaxial pressure triggers a meta-stable crystal structure which stabilizes the pinning of stripe order.
Within statistical errors, the correlation lengths (longitudinal $\xi_\parallel$ and transverse $\xi_\perp$) display no change upon application of uniaxial pressure (Fig.~\ref{fig2}m). Furthermore, we find isotropic correlation lengths $\xi_\parallel\approx \xi_\perp$.\\

\noindent\textbf{Temperature dependence.}
After obtaining the pinned charge-stripe order, we studied its temperature evolution with uniaxial strain released.
With increasing temperature, the charge order peak amplitude $I$ decreases in a $T^{-\eta}$ fashion with $\eta\approx 8/5$ (Fig.~\ref{fig3}(c-f)) up to the highest measured temperature of 120 K.
The correlation length roughly scales with peak amplitude as $I\sim\xi^2$ (see insets of Fig.~\ref{fig3}f,g) and saturates around 20 \AA~in the high temperature limit. Such a scaling behavior was also revealed in La$_{0.165}$Eu$_{0.2}$Sr$_{0.125}$CuO$_4$~\cite{WangPRL2020} and therefore represents a universal characteristic of charge correlation in stripe-ordered cuprates. Finally, we find that the incommensurability $\delta_\parallel$ increases only marginally with temperature and never exceeds 1/4 within our probing window. \\

\noindent\textbf{Discussion} \\ 
To discuss the uniaxial pressure-induced stripe (de)pinning effect, we employ both a phenomenological Landau model~\cite{RobertsonPRB2006,MaestroPRB2006} and a strong-coupling real space picture. Generally, a two-dimensional charge-density wave modulation with wave vectors $\mathbf{Q}_x$ and $\mathbf{Q}_y$ is described by
\begin{equation}
    \delta \rho (\mathbf{r})=\mathrm{Re}(\phi_x e^{i\mathbf{Q}_x\cdot\mathbf{r}}) + \mathrm{Re}(\phi_y e^{i\mathbf{Q}_y\cdot\mathbf{r}}),
\end{equation}
where $\delta \rho (\mathbf{r})$ is the spatial charge modulation and $\phi_i$ with $i={x,y}$ are amplitudes. 
The Landau free energy density for these amplitudes in a tetragonal system is given by:
\begin{align}
    f_0&=\kappa_\parallel(\vert\partial_x\phi_x\vert^2+\vert\partial_y\phi_y\vert^2)+\kappa_\perp(\vert\partial_y\phi_x\vert^2+\vert\partial_x\phi_y\vert^2)\\\nonumber
    &+\alpha(\vert\phi_x\vert^2+\vert\phi_y\vert^2)+\frac{\beta}{2}(\vert\phi_x\vert^2+\vert\phi_y\vert^2)^2 -\gamma\vert\phi_x\vert^2\vert\phi_y\vert^2,\\\nonumber
\end{align}
where the parameters $\alpha$, $\beta$, $\gamma$ describe the homogeneous phase, while $\kappa_\parallel$ and $\kappa_\perp$ link to the longitudinal and transverse stripe order stiffness.
Spontaneous charge order emerges when $\alpha<0$. Checkerboard and stripe orders are found for $\gamma>0$ and $\gamma<0$, respectively. The fourfold symmetry implies that both structures are manifested by reflections at $\mathbf{Q_1}=(\delta_\parallel,0)$ and $\mathbf{Q_2}=(0,\delta_\parallel)$. An orthorhombic distortion with $B_{1g}$ or $B_{2g}$ symmetry adds the following terms to the free energy density~\cite{RobertsonPRB2006}:
\begin{equation}
    f_\mathrm{orth}=-O_{ab}Q_aQ_b\bar{\phi_a}\phi_b+g(O_{ab}iQ_a\bar{\phi_b}\partial_b\phi_a+c.c.),
\end{equation}
where $a,b=x,y$ and $O=h_\parallel\sigma_3$ or $O=h_\perp\sigma_1$ represents the $B_{1g}$ or $B_{2g}$ symmetry breaking strain
with $h_{\parallel(\perp)}$ and $\sigma_j$ being the strain magnitude and the Pauli matrices, respectively.
$g$ is a phenomenological parameter and higher-order terms are neglected.

\begin{figure*}[t]
\center{\includegraphics[width=0.9\textwidth]{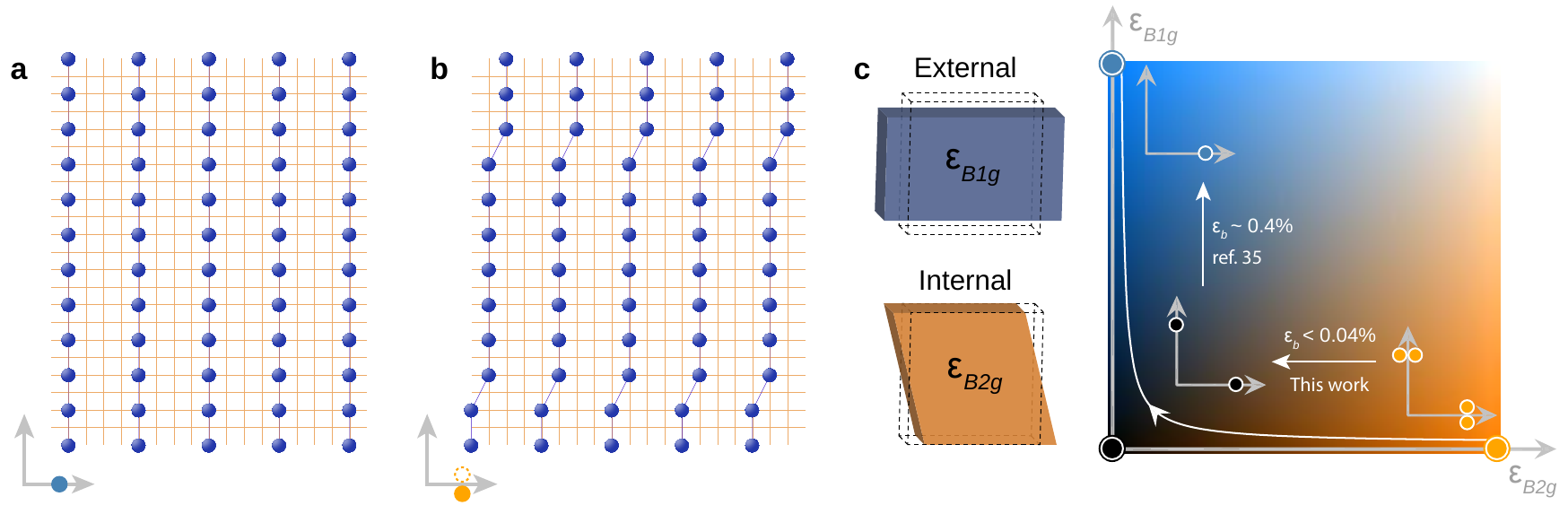}} 
\caption{
\textbf{Stripe order structures and strain phase diagram.}
\textbf{a},\textbf{b} Schematic of half-filled, site-centered stripes.
Orange grid represents the square CuO$_2$ lattice. Spheres illustrate the charge stripes. The in-between antiferromagnetic texture is not shown. \textbf{a} illustrates a statically pinned stripe order. Transverse kink ordering is illustrated in \textbf{b}. The density of aligned kinks along the stripes determines the transverse incommensurability $\delta_\perp$. For unstrained LSCO, the density is about one kink every twenty copper sites. 
Schematic of corresponding diffraction pattern is shown in the bottom-left corner. Dashed open circle indicates the position of charge order peak from the stripe twin domain rotating to the opposite direction.
\textbf{c} Stripe order phase diagram as a function of orthorhombic $B_{1g}$ and $B_{2g}$ strain.
The colour code indicates the parameter space coordinates $(g h_\perp/\kappa_\perp,g h_\parallel/\kappa_\parallel)$, where $g h_i$ with $i={\parallel,\perp}$ represent orthorhombic strain and $\kappa_i$ the stripe stiffness. Blue, black and orange phases represent pinned detwinned, pinned twinned and depinned twinned stripe order, respectively. White curve with arrow indicates schematically the connecting experiment trajectory upon application of uniaxial pressure. The straight white arrows indicate uniaxial pressure (this work and ref.~\cite{Choi20}) applied to connect the different stripe order structures. Schematics on the left-hand side illustrate the two types of orthorhombic lattice distortions.}
\label{fig4}
\end{figure*}

In absence of external strain, LSCO adopts the low-temperature orthorhombic structure with $B_{2g}$ symmetry (see Fig.~\ref{fig4}c). 
In this case, $f_\mathrm{orth}$ gives rise to a small rotation of the ordering vector away from the copper-oxygen bond direction with the new ordering vector being $\mathbf{\tilde{Q}}=\vert\mathbf{Q}\vert(1,gh_\perp/\kappa_\perp)$. Such a transverse incommensuration is indeed observed~\cite{ThampyPRB2014,KimuraPRB2000} (Fig.~\ref{fig2}d,g). Application of external strain along the copper-oxygen bond direction promotes the $B_{1g}$ symmetry breaking terms with magnitude $h_\parallel$. For $h_\parallel\gg h_\perp$, detwinned stripe order pinned to the crystal lattice is expected and recently realized experimentally~\cite{Choi20}. With the modest uniaxial pressure applied in this study, we deem that $h_\parallel\leq h_\perp$. The observation of twinned stripe order with $\delta_\perp \rightarrow 0$ suggests that even modest uniaxial pressure induces a space group change of the crystal structure or reduced orthorhombic distortions. 
A recent numerical calculation using the Hubbard model shows that the stripe rotation is sensitive to the anisotropy of the next-nearest neighbour hopping~\cite{He21}.
This agrees with our finding that modest uniaxial pressure seems to induce an approximate tetragonal crystal field environment.
The evolution from twinned depinned to detwinned pinned stripe order is depicted in Fig.~\ref{fig4}c.

In the real space view~\cite{ZaanenPRB1998,BoschPRB2001,EskesPRB1998}, the stripe incommensuration in unstrained LSCO corresponds to a slanted stripe phase with an average angle of ${\sim}3^{\circ}$ rotated away from the crystal axis. A microscopic picture of the global incommensuration involves domain walls on stripes as elementary excitations of the order parameter~\cite{BoschPRB2001}. Since any charge excitation along the stripe results in an increase in Coulomb energy, transverse excitations are therefore energetically favorable if the curvature energy---reflected by the transverse stiffness $\kappa_\perp$---is small. 
The model thus considers one-dimensional stripes as quantum strings with transverse kinks. It has been shown~\cite{BoschPRB2001} that inter-string coupling leads to a symmetry breaking phase with directional kinks and thus slanted stripes.
Internal or external $B_{1g}$ orthorhombic strain could possibly increase the transverse stripe stiffness $\kappa_\perp$ through an enhanced electron-phonon coupling~\cite{WangSciAdv2021}.
The decrease of the transverse incommensurability $\delta_\perp \propto gh_\perp/\kappa_\perp$ is therefore likely a result of both the reduced $B_{2g}$ orthorhombicity and enhanced stripe stiffness. 
On the other hand, the stripe density is reflected by the longitudinal modulation. The observation of a longitudinal lattice incommensurability ($\delta_\parallel \neq 1/4$) may suggest the presence of stripe disorder (see Fig.~\ref{fig4}a,b and Supplementary Fig.~6). The fact that $\delta_\parallel$ remains essentially unchanged upon application of modest uniaxial pressure indicates that the disorder potential is weakly dependent on pressure. 
Application of larger uniaxial pressure is known~\cite{Choi20} to drive $\delta_\parallel \rightarrow 1/4$.
In the quantum lattice string model, static stripes melt through a transverse kink proliferation~\cite{ZaanenPRB1998,EskesPRB1998}, characterized by the transverse fluctuation magnitude~\cite{KivelsonNat1998}.
Although the energy and time scales of the stripe dynamics~\cite{KivelsonRMP2003,MitranoSciAdv2019} are not directly resolved here, their prominent role is signified by the weak transverse stiffness revealed by our results.
The fact that modest external strain stabilizes the stripe phase suggests that LSCO is in the vicinity of a quantum melting point. The associated quantum fluctuations are likely crucial to the coexistence of superconductivity and stripe phase.\\[2mm]

\noindent\textbf{Methods} \\
\textbf{Samples}. \lsco~single crystals were grown by the floating zone method~\cite{ChangPRB2008}. The superconducting transition temperature is $T_c=27$~K.\\

\noindent\textbf{Resonant inelastic x-ray scattering}. RIXS experiments were carried out at the ADRESS beamline of the Swiss Light Source (SLS) at the Paul Scherrer Institut~\cite{GhiringhelliREVSCIINS2006,StrocovJSYNRAD2010}. The energy resolution---ranging from 124 to 130~meV---is estimated by the full-width-at-half-maximum (FWHM) of the elastic scattering peak from amorphous carbon tape.
To enhance charge scattering, most data were taken using linear vertical ($\sigma$) incident light polarisation with grazing exit scattering geometry. Comparative measurements using horizontal ($\pi$) incident light polarisation have been performed under identical configurations (Supplementary Fig.~3).
Wave vector $\mathbf{Q}=(q_x,q_y,q_z)$ is labeled in reciprocal lattice units (r.l.u.) of $(2\pi/a, 2\pi/b, 2\pi/c)$, where $a=b=3.78$~\AA~and $c=13.2$~\AA~are the lattice parameters of the high-temperature tetragonal unit cell. $q_\parallel$ and $q_\perp$ denote the longitudinal and transverse components of the in-plane momentum $(q_x,q_y)$ in r.l.u., respectively.
Samples were aligned with $a$ and $c$ axes in the horizontal scattering plane and $b$ axis along the vertical direction. The scattering angle was fixed at $2\theta = 130^{\circ}$ (see Fig.~\ref{fig1}b).
In-plane momentum is set by varying the $\omega$ and $\phi$ angles (see Fig.~\ref{fig1}b). \\

\noindent\textbf{Uniaxial strain application}. To apply uniaxial strain, we adapted a tool previously used to cleave crystals for angle-resolved photoemission spectroscopy experiments~\cite{cleaver}. 
For the RIXS measurements, our LSCO crystals were cleaved using a top-post. Uniaxial pressure was applied at low temperature (${\sim}28$~K) along a copper-oxygen bond direction through an \textit{in-situ} operational screw mechanism (see Fig.~\ref{fig1}a). \\

\noindent\textbf{Analysis of RIXS data}. RIXS intensities are normalised to the weight of $dd$ excitations~\cite{WangPRL2020}.
Elastic intensity is extracted by integrating the spectral weight around zero energy loss within $\pm$FWHM energy window. We have also analysed the data by defining the area of the fitted Gaussian elastic line as the elastic intensity. The two analysis methodologies yield consistent conclusions (Supplementary Figs.~7,8).
Correlation lengths are defined as the inverse half-width-at-half-maximum. \\

\noindent\textbf{X-ray diffraction}. XRD measurements were performed at 300~K where uniaxial strain was applied.
Since the elastic constants of the sample~\cite{MiglioriPRB1990} and materials used for the strain cell (type 316 stainless steel~\cite{LedbetterJAP1981} and copper~\cite{OvertonPR1998}) change only slightly below 300~K, the maximum strain values applied in the RIXS and XRD experiments are comparable.\\[1mm]

\noindent\textbf{Acknowledgments}\\
We thank Ke-Jin Zhou for insightful discussions. Q.W., K.v.A., M.H., W.Z., T.C.A., T.S., and J.C. acknowledge support by the Swiss National Science Foundation through Grant Numbers BSSGI0\_155873, 200021\_188564 and 200021\_178867. K.v.A. is grateful for the support from the FAN Research Talent Development Fund-UZH Alumni.
Q.W. and K.v.A. thank the Forschungskredit of the University of Zurich, under grant numbers [FK-20-128] and [FK-21-105]. 
T.C.A. acknowledges funding from the European Union’s Horizon 2020 research and innovation programme under the Marie Skłodowska-Curie grant agreement No. 701647 (PSI-FELLOW-II-3i program).
Y.S. thanks the Chalmers Area of Advances-Materials Science and the Swedish Research Council (VR) with the starting grant (Dnr. 2017-05078) for funding. N.B.C. thanks the Danish Agency for Science, Technology, and Innovation for funding the instrument center DanScatt and acknowledges support from the Q-MAT ESS Lighthouse initiative. M.M. is funded by the Swedish Research Council (VR) through a neutron project grant (Dnr. 2016-06955) as well as the KTH Materials Platform.\\

\vspace{2mm}
\noindent\textbf{Authors contributions}\\
Q.W. and J.~Chang conceived the project. T.K., N.M., and M.O. grew and characterised the LSCO single crystals. Q.W., J.~Choi, D.B., M.M., and J.~Chang designed the uniaxial strain device. Q.W., K.v.A., S.M., M.H., D.M., W.Z., T.C.A., Y.S., T.S., and J.~Chang carried out the RIXS experiments. Q.W., H.W., and J.Z. perfomed the XRD measurements. Q.W. and J.~Chang analysed the data with support from J.K., N.B.C., M.H.F., and M.J.. Q.W. and J.~Chang wrote the manuscript with input from all authors.

\vspace{2mm}
\noindent\textbf{Competing interests}\\
The authors declare no competing interests.

\vspace{2mm}
\noindent\textbf{Data and materials availability}\\ 
All data that support the findings of this study are available from the corresponding authors upon reasonable request.\\

\clearpage


\begin{thebibliography}{49}%
	\makeatletter
	\providecommand \@ifxundefined [1]{%
		\@ifx{#1\undefined}
	}%
	\providecommand \@ifnum [1]{%
		\ifnum #1\expandafter \@firstoftwo
		\else \expandafter \@secondoftwo
		\fi
	}%
	\providecommand \@ifx [1]{%
		\ifx #1\expandafter \@firstoftwo
		\else \expandafter \@secondoftwo
		\fi
	}%
	\providecommand \natexlab [1]{#1}%
	\providecommand \enquote  [1]{``#1''}%
	\providecommand \bibnamefont  [1]{#1}%
	\providecommand \bibfnamefont [1]{#1}%
	\providecommand \citenamefont [1]{#1}%
	\providecommand \href@noop [0]{\@secondoftwo}%
	\providecommand \href [0]{\begingroup \@sanitize@url \@href}%
	\providecommand \@href[1]{\@@startlink{#1}\@@href}%
	\providecommand \@@href[1]{\endgroup#1\@@endlink}%
	\providecommand \@sanitize@url [0]{\catcode `\\12\catcode `\$12\catcode
		`\&12\catcode `\#12\catcode `\^12\catcode `\_12\catcode `\%12\relax}%
	\providecommand \@@startlink[1]{}%
	\providecommand \@@endlink[0]{}%
	\providecommand \url  [0]{\begingroup\@sanitize@url \@url }%
	\providecommand \@url [1]{\endgroup\@href {#1}{\urlprefix }}%
	\providecommand \urlprefix  [0]{URL }%
	\providecommand \Eprint [0]{\href }%
	\providecommand \doibase [0]{http://dx.doi.org/}%
	\providecommand \selectlanguage [0]{\@gobble}%
	\providecommand \bibinfo  [0]{\@secondoftwo}%
	\providecommand \bibfield  [0]{\@secondoftwo}%
	\providecommand \translation [1]{[#1]}%
	\providecommand \BibitemOpen [0]{}%
	\providecommand \bibitemStop [0]{}%
	\providecommand \bibitemNoStop [0]{.\EOS\space}%
	\providecommand \EOS [0]{\spacefactor3000\relax}%
	\providecommand \BibitemShut  [1]{\csname bibitem#1\endcsname}%
	\let\auto@bib@innerbib\@empty
	\bibitem [{\citenamefont {Machida}(1989)}]{MachidaPhysC1989}%
	\BibitemOpen
	\bibfield  {author} {\bibinfo {author} {\bibfnamefont {K.}~\bibnamefont
			{Machida}},\ }\bibfield  {title} {\enquote {\bibinfo {title} {Magnetism in
				{${\mathrm{La}}_{2}{\mathrm{CuO}}_{4}$} based compounds},}\ }\href {\doibase
		10.1016/0921-4534(89)90316-x} {\bibfield  {journal} {\bibinfo  {journal}
			{Physica C}\ }\textbf {\bibinfo {volume} {158}},\ \bibinfo {pages} {192--196}
		(\bibinfo {year} {1989})}\BibitemShut {NoStop}%
	\bibitem [{\citenamefont {Zaanen}\ and\ \citenamefont
		{Gunnarsson}(1989)}]{ZaanenPRB1989}%
	\BibitemOpen
	\bibfield  {author} {\bibinfo {author} {\bibfnamefont {J.}~\bibnamefont
			{Zaanen}}\ and\ \bibinfo {author} {\bibfnamefont {O.}~\bibnamefont
			{Gunnarsson}},\ }\bibfield  {title} {\enquote {\bibinfo {title} {Charged
				magnetic domain lines and the magnetism of high-${T}_{c}$ oxides},}\ }\href
	{\doibase 10.1103/PhysRevB.40.7391} {\bibfield  {journal} {\bibinfo
			{journal} {Phys. Rev. B}\ }\textbf {\bibinfo {volume} {40}},\ \bibinfo
		{pages} {7391--7394} (\bibinfo {year} {1989})}\BibitemShut {NoStop}%
	\bibitem [{\citenamefont {Emery}\ and\ \citenamefont
		{Kivelson}(1993)}]{EmeryPhysC1993}%
	\BibitemOpen
	\bibfield  {author} {\bibinfo {author} {\bibfnamefont {V.}~\bibnamefont
			{Emery}}\ and\ \bibinfo {author} {\bibfnamefont {S.}~\bibnamefont
			{Kivelson}},\ }\bibfield  {title} {\enquote {\bibinfo {title} {Frustrated
				electronic phase separation and high-temperature superconductors},}\ }\href
	{\doibase https://doi.org/10.1016/0921-4534(93)90581-A} {\bibfield  {journal}
		{\bibinfo  {journal} {Physica C: Superconductivity}\ }\textbf {\bibinfo
			{volume} {209}},\ \bibinfo {pages} {597--621} (\bibinfo {year}
		{1993})}\BibitemShut {NoStop}%
	\bibitem [{\citenamefont {Castellani}\ \emph {et~al.}(1995)\citenamefont
		{Castellani}, \citenamefont {Di~Castro},\ and\ \citenamefont
		{Grilli}}]{CastellaniPRL1995}%
	\BibitemOpen
	\bibfield  {author} {\bibinfo {author} {\bibfnamefont {C.}~\bibnamefont
			{Castellani}}, \bibinfo {author} {\bibfnamefont {C.}~\bibnamefont
			{Di~Castro}}, \ and\ \bibinfo {author} {\bibfnamefont {M.}~\bibnamefont
			{Grilli}},\ }\bibfield  {title} {\enquote {\bibinfo {title} {Singular
				quasiparticle scattering in the proximity of charge instabilities},}\ }\href
	{\doibase 10.1103/PhysRevLett.75.4650} {\bibfield  {journal} {\bibinfo
			{journal} {Phys. Rev. Lett.}\ }\textbf {\bibinfo {volume} {75}},\ \bibinfo
		{pages} {4650--4653} (\bibinfo {year} {1995})}\BibitemShut {NoStop}%
	\bibitem [{\citenamefont {Hellberg}\ and\ \citenamefont
		{Manousakis}(1997)}]{HellbergPRL1997}%
	\BibitemOpen
	\bibfield  {author} {\bibinfo {author} {\bibfnamefont {C.~S.}\ \bibnamefont
			{Hellberg}}\ and\ \bibinfo {author} {\bibfnamefont {E.}~\bibnamefont
			{Manousakis}},\ }\bibfield  {title} {\enquote {\bibinfo {title} {Phase
				separation at all interaction strengths in the {$t$-$J$} model},}\ }\href
	{\doibase 10.1103/PhysRevLett.78.4609} {\bibfield  {journal} {\bibinfo
			{journal} {Phys. Rev. Lett.}\ }\textbf {\bibinfo {volume} {78}},\ \bibinfo
		{pages} {4609--4612} (\bibinfo {year} {1997})}\BibitemShut {NoStop}%
	\bibitem [{\citenamefont {Seibold}\ \emph {et~al.}(1998)\citenamefont
		{Seibold}, \citenamefont {Sigmund},\ and\ \citenamefont
		{Hizhnyakov}}]{SeiboldPRB1998}%
	\BibitemOpen
	\bibfield  {author} {\bibinfo {author} {\bibfnamefont {G.}~\bibnamefont
			{Seibold}}, \bibinfo {author} {\bibfnamefont {E.}~\bibnamefont {Sigmund}}, \
		and\ \bibinfo {author} {\bibfnamefont {V.}~\bibnamefont {Hizhnyakov}},\
	}\bibfield  {title} {\enquote {\bibinfo {title} {Unrestricted slave-boson
				mean-field approximation for the two-dimensional {Hubbard} model},}\ }\href
	{\doibase 10.1103/PhysRevB.57.6937} {\bibfield  {journal} {\bibinfo
			{journal} {Phys. Rev. B}\ }\textbf {\bibinfo {volume} {57}},\ \bibinfo
		{pages} {6937--6942} (\bibinfo {year} {1998})}\BibitemShut {NoStop}%
	\bibitem [{\citenamefont {Huang}\ \emph {et~al.}(2017)\citenamefont {Huang},
		\citenamefont {Mendl}, \citenamefont {Liu}, \citenamefont {Johnston},
		\citenamefont {Jiang}, \citenamefont {Moritz},\ and\ \citenamefont
		{Devereaux}}]{HuangSci2017}%
	\BibitemOpen
	\bibfield  {author} {\bibinfo {author} {\bibfnamefont {E.~W.}\ \bibnamefont
			{Huang}}, \bibinfo {author} {\bibfnamefont {C.~B.}\ \bibnamefont {Mendl}},
		\bibinfo {author} {\bibfnamefont {S.}~\bibnamefont {Liu}}, \bibinfo {author}
		{\bibfnamefont {S.}~\bibnamefont {Johnston}}, \bibinfo {author}
		{\bibfnamefont {H.-C.}\ \bibnamefont {Jiang}}, \bibinfo {author}
		{\bibfnamefont {B.}~\bibnamefont {Moritz}}, \ and\ \bibinfo {author}
		{\bibfnamefont {T.~P.}\ \bibnamefont {Devereaux}},\ }\bibfield  {title}
	{\enquote {\bibinfo {title} {Numerical evidence of fluctuating stripes in the
				normal state of high-${T}_{c}$ cuprate superconductors},}\ }\href {\doibase
		10.1126/science.aak9546} {\bibfield  {journal} {\bibinfo  {journal}
			{Science}\ }\textbf {\bibinfo {volume} {358}},\ \bibinfo {pages} {1161--1164}
		(\bibinfo {year} {2017})}\BibitemShut {NoStop}%
	\bibitem [{\citenamefont {Zheng}\ \emph {et~al.}(2017)\citenamefont {Zheng},
		\citenamefont {Chung}, \citenamefont {Corboz}, \citenamefont {Ehlers},
		\citenamefont {Qin}, \citenamefont {Noack}, \citenamefont {Shi},
		\citenamefont {White}, \citenamefont {Zhang},\ and\ \citenamefont
		{Chan}}]{ZhengSci2017}%
	\BibitemOpen
	\bibfield  {author} {\bibinfo {author} {\bibfnamefont {B.-X.}\ \bibnamefont
			{Zheng}}, \bibinfo {author} {\bibfnamefont {C.-M.}\ \bibnamefont {Chung}},
		\bibinfo {author} {\bibfnamefont {P.}~\bibnamefont {Corboz}}, \bibinfo
		{author} {\bibfnamefont {G.}~\bibnamefont {Ehlers}}, \bibinfo {author}
		{\bibfnamefont {M.-P.}\ \bibnamefont {Qin}}, \bibinfo {author} {\bibfnamefont
			{R.~M.}\ \bibnamefont {Noack}}, \bibinfo {author} {\bibfnamefont
			{H.}~\bibnamefont {Shi}}, \bibinfo {author} {\bibfnamefont {S.~R.}\
			\bibnamefont {White}}, \bibinfo {author} {\bibfnamefont {S.}~\bibnamefont
			{Zhang}}, \ and\ \bibinfo {author} {\bibfnamefont {G.~K.-L.}\ \bibnamefont
			{Chan}},\ }\bibfield  {title} {\enquote {\bibinfo {title} {Stripe order in
				the underdoped region of the two-dimensional {Hubbard} model},}\ }\href
	{\doibase 10.1126/science.aam7127} {\bibfield  {journal} {\bibinfo  {journal}
			{Science}\ }\textbf {\bibinfo {volume} {358}},\ \bibinfo {pages} {1155--1160}
		(\bibinfo {year} {2017})}\BibitemShut {NoStop}%
	\bibitem [{\citenamefont {Tranquada}\ \emph {et~al.}(1995)\citenamefont
		{Tranquada}, \citenamefont {Sternlieb}, \citenamefont {Axe}, \citenamefont
		{Nakamura},\ and\ \citenamefont {Uchida}}]{TranquadaNat1995}%
	\BibitemOpen
	\bibfield  {author} {\bibinfo {author} {\bibfnamefont {J.~M.}\ \bibnamefont
			{Tranquada}}, \bibinfo {author} {\bibfnamefont {B.~J.}\ \bibnamefont
			{Sternlieb}}, \bibinfo {author} {\bibfnamefont {J.~D.}\ \bibnamefont {Axe}},
		\bibinfo {author} {\bibfnamefont {Y.}~\bibnamefont {Nakamura}}, \ and\
		\bibinfo {author} {\bibfnamefont {S.}~\bibnamefont {Uchida}},\ }\bibfield
	{title} {\enquote {\bibinfo {title} {Evidence for stripe correlations of
				spins and holes in copper oxide superconductors},}\ }\href {\doibase
		10.1038/375561a0} {\bibfield  {journal} {\bibinfo  {journal} {Nature
				(London)}\ }\textbf {\bibinfo {volume} {375}},\ \bibinfo {pages} {561--563}
		(\bibinfo {year} {1995})}\BibitemShut {NoStop}%
	\bibitem [{\citenamefont {Emery}\ and\ \citenamefont
		{Kivelson}(1994)}]{EmeryPhysC1994}%
	\BibitemOpen
	\bibfield  {author} {\bibinfo {author} {\bibfnamefont {V.~J.}\ \bibnamefont
			{Emery}}\ and\ \bibinfo {author} {\bibfnamefont {S.~A.}\ \bibnamefont
			{Kivelson}},\ }\bibfield  {title} {\enquote {\bibinfo {title} {Collective
				charge transport in high temperature superconductors},}\ }\href {\doibase
		10.1016/0921-4534(94)91345-5} {\bibfield  {journal} {\bibinfo  {journal}
			{Physica C: Superconductivity}\ }\textbf {\bibinfo {volume} {235-240}},\
		\bibinfo {pages} {189--192} (\bibinfo {year} {1994})}\BibitemShut {NoStop}%
	\bibitem [{\citenamefont {Chang}\ \emph {et~al.}(2008)\citenamefont {Chang},
		\citenamefont {Niedermayer}, \citenamefont {Gilardi}, \citenamefont
		{Christensen}, \citenamefont {R\o{}nnow}, \citenamefont {McMorrow},
		\citenamefont {Ay}, \citenamefont {Stahn}, \citenamefont {Sobolev},
		\citenamefont {Hiess}, \citenamefont {Pailhes}, \citenamefont {Baines},
		\citenamefont {Momono}, \citenamefont {Oda}, \citenamefont {Ido},\ and\
		\citenamefont {Mesot}}]{ChangPRB2008}%
	\BibitemOpen
	\bibfield  {author} {\bibinfo {author} {\bibfnamefont {J.}~\bibnamefont
			{Chang}}, \bibinfo {author} {\bibfnamefont {C.}~\bibnamefont {Niedermayer}},
		\bibinfo {author} {\bibfnamefont {R.}~\bibnamefont {Gilardi}}, \bibinfo
		{author} {\bibfnamefont {N.~B.}\ \bibnamefont {Christensen}}, \bibinfo
		{author} {\bibfnamefont {H.~M.}\ \bibnamefont {R\o{}nnow}}, \bibinfo {author}
		{\bibfnamefont {D.~F.}\ \bibnamefont {McMorrow}}, \bibinfo {author}
		{\bibfnamefont {M.}~\bibnamefont {Ay}}, \bibinfo {author} {\bibfnamefont
			{J.}~\bibnamefont {Stahn}}, \bibinfo {author} {\bibfnamefont
			{O.}~\bibnamefont {Sobolev}}, \bibinfo {author} {\bibfnamefont
			{A.}~\bibnamefont {Hiess}}, \bibinfo {author} {\bibfnamefont
			{S.}~\bibnamefont {Pailhes}}, \bibinfo {author} {\bibfnamefont
			{C.}~\bibnamefont {Baines}}, \bibinfo {author} {\bibfnamefont
			{N.}~\bibnamefont {Momono}}, \bibinfo {author} {\bibfnamefont
			{M.}~\bibnamefont {Oda}}, \bibinfo {author} {\bibfnamefont {M.}~\bibnamefont
			{Ido}}, \ and\ \bibinfo {author} {\bibfnamefont {J.}~\bibnamefont {Mesot}},\
	}\bibfield  {title} {\enquote {\bibinfo {title} {Tuning competing orders in
				{La}$_{2-x}${Sr}$_{x}${CuO}$_{4}$ cuprate superconductors by the application
				of an external magnetic field},}\ }\href {\doibase
		10.1103/PhysRevB.78.104525} {\bibfield  {journal} {\bibinfo  {journal} {Phys.
				Rev. B}\ }\textbf {\bibinfo {volume} {78}},\ \bibinfo {pages} {104525}
		(\bibinfo {year} {2008})}\BibitemShut {NoStop}%
	\bibitem [{\citenamefont {H\"ucker}\ \emph {et~al.}(2011)\citenamefont
		{H\"ucker}, \citenamefont {v.~Zimmermann}, \citenamefont {Gu}, \citenamefont
		{Xu}, \citenamefont {Wen}, \citenamefont {Xu}, \citenamefont {Kang},
		\citenamefont {Zheludev},\ and\ \citenamefont {Tranquada}}]{HuckerPRB2011}%
	\BibitemOpen
	\bibfield  {author} {\bibinfo {author} {\bibfnamefont {M.}~\bibnamefont
			{H\"ucker}}, \bibinfo {author} {\bibfnamefont {M.}~\bibnamefont
			{v.~Zimmermann}}, \bibinfo {author} {\bibfnamefont {G.~D.}\ \bibnamefont
			{Gu}}, \bibinfo {author} {\bibfnamefont {Z.~J.}\ \bibnamefont {Xu}}, \bibinfo
		{author} {\bibfnamefont {J.~S.}\ \bibnamefont {Wen}}, \bibinfo {author}
		{\bibfnamefont {G.}~\bibnamefont {Xu}}, \bibinfo {author} {\bibfnamefont
			{H.~J.}\ \bibnamefont {Kang}}, \bibinfo {author} {\bibfnamefont
			{A.}~\bibnamefont {Zheludev}}, \ and\ \bibinfo {author} {\bibfnamefont
			{J.~M.}\ \bibnamefont {Tranquada}},\ }\bibfield  {title} {\enquote {\bibinfo
			{title} {Stripe order in superconducting {La}$_{2-x}${Ba}$_{x}${CuO}$_{4}$
				($0.095\leqslant x \leqslant 0.155$)},}\ }\href {\doibase
		10.1103/PhysRevB.83.104506} {\bibfield  {journal} {\bibinfo  {journal} {Phys.
				Rev. B}\ }\textbf {\bibinfo {volume} {83}},\ \bibinfo {pages} {104506}
		(\bibinfo {year} {2011})}\BibitemShut {NoStop}%
	\bibitem [{\citenamefont {Kivelson}\ \emph {et~al.}(1998)\citenamefont
		{Kivelson}, \citenamefont {Fradkin},\ and\ \citenamefont
		{Emery}}]{KivelsonNat1998}%
	\BibitemOpen
	\bibfield  {author} {\bibinfo {author} {\bibfnamefont {S.~A.}\ \bibnamefont
			{Kivelson}}, \bibinfo {author} {\bibfnamefont {E.}~\bibnamefont {Fradkin}}, \
		and\ \bibinfo {author} {\bibfnamefont {V.~J.}\ \bibnamefont {Emery}},\
	}\bibfield  {title} {\enquote {\bibinfo {title} {Electronic liquid-crystal
				phases of a doped {Mott} insulator},}\ }\href {\doibase 10.1038/31177}
	{\bibfield  {journal} {\bibinfo  {journal} {Nature}\ }\textbf {\bibinfo
			{volume} {393}},\ \bibinfo {pages} {550--553} (\bibinfo {year}
		{1998})}\BibitemShut {NoStop}%
	\bibitem [{\citenamefont {Kivelson}\ \emph {et~al.}(2003)\citenamefont
		{Kivelson}, \citenamefont {Bindloss}, \citenamefont {Fradkin}, \citenamefont
		{Oganesyan}, \citenamefont {Tranquada}, \citenamefont {Kapitulnik},\ and\
		\citenamefont {Howald}}]{KivelsonRMP2003}%
	\BibitemOpen
	\bibfield  {author} {\bibinfo {author} {\bibfnamefont {S.~A.}\ \bibnamefont
			{Kivelson}}, \bibinfo {author} {\bibfnamefont {I.~P.}\ \bibnamefont
			{Bindloss}}, \bibinfo {author} {\bibfnamefont {E.}~\bibnamefont {Fradkin}},
		\bibinfo {author} {\bibfnamefont {V.}~\bibnamefont {Oganesyan}}, \bibinfo
		{author} {\bibfnamefont {J.~M.}\ \bibnamefont {Tranquada}}, \bibinfo {author}
		{\bibfnamefont {A.}~\bibnamefont {Kapitulnik}}, \ and\ \bibinfo {author}
		{\bibfnamefont {C.}~\bibnamefont {Howald}},\ }\bibfield  {title} {\enquote
		{\bibinfo {title} {How to detect fluctuating stripes in the high-temperature
				superconductors},}\ }\href {\doibase 10.1103/RevModPhys.75.1201} {\bibfield
		{journal} {\bibinfo  {journal} {Rev. Mod. Phys.}\ }\textbf {\bibinfo {volume}
			{75}},\ \bibinfo {pages} {1201--1241} (\bibinfo {year} {2003})}\BibitemShut
	{NoStop}%
	\bibitem [{\citenamefont {Scalapino}(2012)}]{ScalapinoRMP2012}%
	\BibitemOpen
	\bibfield  {author} {\bibinfo {author} {\bibfnamefont {D.~J.}\ \bibnamefont
			{Scalapino}},\ }\bibfield  {title} {\enquote {\bibinfo {title} {A common
				thread: The pairing interaction for unconventional superconductors},}\ }\href
	{\doibase 10.1103/RevModPhys.84.1383} {\bibfield  {journal} {\bibinfo
			{journal} {Rev. Mod. Phys.}\ }\textbf {\bibinfo {volume} {84}},\ \bibinfo
		{pages} {1383--1417} (\bibinfo {year} {2012})}\BibitemShut {NoStop}%
	\bibitem [{\citenamefont {Tranquada}\ \emph {et~al.}(1994)\citenamefont
		{Tranquada}, \citenamefont {Buttrey}, \citenamefont {Sachan},\ and\
		\citenamefont {Lorenzo}}]{TranquadaPRL1994}%
	\BibitemOpen
	\bibfield  {author} {\bibinfo {author} {\bibfnamefont {J.~M.}\ \bibnamefont
			{Tranquada}}, \bibinfo {author} {\bibfnamefont {D.~J.}\ \bibnamefont
			{Buttrey}}, \bibinfo {author} {\bibfnamefont {V.}~\bibnamefont {Sachan}}, \
		and\ \bibinfo {author} {\bibfnamefont {J.~E.}\ \bibnamefont {Lorenzo}},\
	}\bibfield  {title} {\enquote {\bibinfo {title} {Simultaneous ordering of
				holes and spins in {La}$_{2}${NiO}$_{4.125}$},}\ }\href {\doibase
		10.1103/PhysRevLett.73.1003} {\bibfield  {journal} {\bibinfo  {journal}
			{Phys. Rev. Lett.}\ }\textbf {\bibinfo {volume} {73}},\ \bibinfo {pages}
		{1003--1006} (\bibinfo {year} {1994})}\BibitemShut {NoStop}%
	\bibitem [{\citenamefont {Shimomura}\ \emph {et~al.}(1999)\citenamefont
		{Shimomura}, \citenamefont {Wakabayashi}, \citenamefont {Kuwahara},\ and\
		\citenamefont {Tokura}}]{ShimomuraPRL1999}%
	\BibitemOpen
	\bibfield  {author} {\bibinfo {author} {\bibfnamefont {S.}~\bibnamefont
			{Shimomura}}, \bibinfo {author} {\bibfnamefont {N.}~\bibnamefont
			{Wakabayashi}}, \bibinfo {author} {\bibfnamefont {H.}~\bibnamefont
			{Kuwahara}}, \ and\ \bibinfo {author} {\bibfnamefont {Y.}~\bibnamefont
			{Tokura}},\ }\bibfield  {title} {\enquote {\bibinfo {title} {X-ray diffuse
				scattering due to polarons in a colossal magnetoresistive manganite},}\
	}\href {\doibase 10.1103/PhysRevLett.83.4389} {\bibfield  {journal} {\bibinfo
			{journal} {Phys. Rev. Lett.}\ }\textbf {\bibinfo {volume} {83}},\ \bibinfo
		{pages} {4389--4392} (\bibinfo {year} {1999})}\BibitemShut {NoStop}%
	\bibitem [{\citenamefont {Vasiliu-Doloc}\ \emph {et~al.}(1999)\citenamefont
		{Vasiliu-Doloc}, \citenamefont {Rosenkranz}, \citenamefont {Osborn},
		\citenamefont {Sinha}, \citenamefont {Lynn}, \citenamefont {Mesot},
		\citenamefont {Seeck}, \citenamefont {Preosti}, \citenamefont {Fedro},\ and\
		\citenamefont {Mitchell}}]{Vasiliu-DolocPRL1999}%
	\BibitemOpen
	\bibfield  {author} {\bibinfo {author} {\bibfnamefont {L.}~\bibnamefont
			{Vasiliu-Doloc}}, \bibinfo {author} {\bibfnamefont {S.}~\bibnamefont
			{Rosenkranz}}, \bibinfo {author} {\bibfnamefont {R.}~\bibnamefont {Osborn}},
		\bibinfo {author} {\bibfnamefont {S.~K.}\ \bibnamefont {Sinha}}, \bibinfo
		{author} {\bibfnamefont {J.~W.}\ \bibnamefont {Lynn}}, \bibinfo {author}
		{\bibfnamefont {J.}~\bibnamefont {Mesot}}, \bibinfo {author} {\bibfnamefont
			{O.~H.}\ \bibnamefont {Seeck}}, \bibinfo {author} {\bibfnamefont
			{G.}~\bibnamefont {Preosti}}, \bibinfo {author} {\bibfnamefont {A.~J.}\
			\bibnamefont {Fedro}}, \ and\ \bibinfo {author} {\bibfnamefont {J.~F.}\
			\bibnamefont {Mitchell}},\ }\bibfield  {title} {\enquote {\bibinfo {title}
			{Charge melting and polaron collapse in
				{${\mathrm{La}}_{1.2}{\mathrm{Sr}}_{1.8}{\mathrm{Mn}}_{2}{O}_{7}$}},}\ }\href
	{\doibase 10.1103/PhysRevLett.83.4393} {\bibfield  {journal} {\bibinfo
			{journal} {Phys. Rev. Lett.}\ }\textbf {\bibinfo {volume} {83}},\ \bibinfo
		{pages} {4393--4396} (\bibinfo {year} {1999})}\BibitemShut {NoStop}%
	\bibitem [{\citenamefont {Bosch}\ \emph {et~al.}(2001)\citenamefont {Bosch},
		\citenamefont {van Saarloos},\ and\ \citenamefont {Zaanen}}]{BoschPRB2001}%
	\BibitemOpen
	\bibfield  {author} {\bibinfo {author} {\bibfnamefont {M.}~\bibnamefont
			{Bosch}}, \bibinfo {author} {\bibfnamefont {W.}~\bibnamefont {van Saarloos}},
		\ and\ \bibinfo {author} {\bibfnamefont {J.}~\bibnamefont {Zaanen}},\
	}\bibfield  {title} {\enquote {\bibinfo {title} {Shifting bragg peaks of
				cuprate stripes as possible indications for fractionally charged kinks},}\
	}\href {\doibase 10.1103/PhysRevB.63.092501} {\bibfield  {journal} {\bibinfo
			{journal} {Phys. Rev. B}\ }\textbf {\bibinfo {volume} {63}},\ \bibinfo
		{pages} {092501} (\bibinfo {year} {2001})}\BibitemShut {NoStop}%
	\bibitem [{\citenamefont {Zaanen}\ \emph {et~al.}(1998)\citenamefont {Zaanen},
		\citenamefont {Osman},\ and\ \citenamefont {van Saarloos}}]{ZaanenPRB1998}%
	\BibitemOpen
	\bibfield  {author} {\bibinfo {author} {\bibfnamefont {J.}~\bibnamefont
			{Zaanen}}, \bibinfo {author} {\bibfnamefont {O.~Y.}\ \bibnamefont {Osman}}, \
		and\ \bibinfo {author} {\bibfnamefont {W.}~\bibnamefont {van Saarloos}},\
	}\bibfield  {title} {\enquote {\bibinfo {title} {Metallic stripes: Separation
				of spin, charge, and string fluctuation},}\ }\href {\doibase
		10.1103/PhysRevB.58.R11868} {\bibfield  {journal} {\bibinfo  {journal} {Phys.
				Rev. B}\ }\textbf {\bibinfo {volume} {58}},\ \bibinfo {pages} {R11868(R)}
		(\bibinfo {year} {1998})}\BibitemShut {NoStop}%
	\bibitem [{\citenamefont {Mitrano}\ \emph {et~al.}(2019)\citenamefont
		{Mitrano}, \citenamefont {Lee}, \citenamefont {Husain}, \citenamefont
		{Delacretaz}, \citenamefont {Zhu}, \citenamefont {de~la Pe{\~n}a~Munoz},
		\citenamefont {Sun}, \citenamefont {Joe}, \citenamefont {Reid}, \citenamefont
		{Wandel}, \citenamefont {Coslovich}, \citenamefont {Schlotter}, \citenamefont
		{van Driel}, \citenamefont {Schneeloch}, \citenamefont {Gu}, \citenamefont
		{Hartnoll}, \citenamefont {Goldenfeld},\ and\ \citenamefont
		{Abbamonte}}]{MitranoSciAdv2019}%
	\BibitemOpen
	\bibfield  {author} {\bibinfo {author} {\bibfnamefont {M.}~\bibnamefont
			{Mitrano}}, \bibinfo {author} {\bibfnamefont {S.}~\bibnamefont {Lee}},
		\bibinfo {author} {\bibfnamefont {A.~A.}\ \bibnamefont {Husain}}, \bibinfo
		{author} {\bibfnamefont {L.}~\bibnamefont {Delacretaz}}, \bibinfo {author}
		{\bibfnamefont {M.}~\bibnamefont {Zhu}}, \bibinfo {author} {\bibfnamefont
			{G.}~\bibnamefont {de~la Pe{\~n}a~Munoz}}, \bibinfo {author} {\bibfnamefont
			{S.~X.-L.}\ \bibnamefont {Sun}}, \bibinfo {author} {\bibfnamefont {Y.~I.}\
			\bibnamefont {Joe}}, \bibinfo {author} {\bibfnamefont {A.~H.}\ \bibnamefont
			{Reid}}, \bibinfo {author} {\bibfnamefont {S.~F.}\ \bibnamefont {Wandel}},
		\bibinfo {author} {\bibfnamefont {G.}~\bibnamefont {Coslovich}}, \bibinfo
		{author} {\bibfnamefont {W.}~\bibnamefont {Schlotter}}, \bibinfo {author}
		{\bibfnamefont {T.}~\bibnamefont {van Driel}}, \bibinfo {author}
		{\bibfnamefont {J.}~\bibnamefont {Schneeloch}}, \bibinfo {author}
		{\bibfnamefont {G.~D.}\ \bibnamefont {Gu}}, \bibinfo {author} {\bibfnamefont
			{S.}~\bibnamefont {Hartnoll}}, \bibinfo {author} {\bibfnamefont
			{N.}~\bibnamefont {Goldenfeld}}, \ and\ \bibinfo {author} {\bibfnamefont
			{P.}~\bibnamefont {Abbamonte}},\ }\bibfield  {title} {\enquote {\bibinfo
			{title} {Ultrafast time-resolved x-ray scattering reveals diffusive charge
				order dynamics in {La}$_{2-x}${Ba}$_{x}${CuO}$_{4}$},}\ }\href {\doibase
		10.1126/sciadv.aax3346} {\bibfield  {journal} {\bibinfo  {journal} {Sci.
				Adv.}\ }\textbf {\bibinfo {volume} {5}},\ \bibinfo {pages} {eaax3346}
		(\bibinfo {year} {2019})}\BibitemShut {NoStop}%
	\bibitem [{\citenamefont {Croft}\ \emph {et~al.}(2014)\citenamefont {Croft},
		\citenamefont {Lester}, \citenamefont {Senn}, \citenamefont {Bombardi},\ and\
		\citenamefont {Hayden}}]{CroftPRB2014}%
	\BibitemOpen
	\bibfield  {author} {\bibinfo {author} {\bibfnamefont {T.~P.}\ \bibnamefont
			{Croft}}, \bibinfo {author} {\bibfnamefont {C.}~\bibnamefont {Lester}},
		\bibinfo {author} {\bibfnamefont {M.~S.}\ \bibnamefont {Senn}}, \bibinfo
		{author} {\bibfnamefont {A.}~\bibnamefont {Bombardi}}, \ and\ \bibinfo
		{author} {\bibfnamefont {S.~M.}\ \bibnamefont {Hayden}},\ }\bibfield  {title}
	{\enquote {\bibinfo {title} {Charge density wave fluctuations in
				{La}$_{2-x}${Sr}$_{x}${CuO}$_{4}$ and their competition with
				superconductivity},}\ }\href {\doibase 10.1103/PhysRevB.89.224513} {\bibfield
		{journal} {\bibinfo  {journal} {Phys. Rev. B}\ }\textbf {\bibinfo {volume}
			{89}},\ \bibinfo {pages} {224513} (\bibinfo {year} {2014})}\BibitemShut
	{NoStop}%
	\bibitem [{\citenamefont {Christensen}\ \emph {et~al.}()\citenamefont
		{Christensen}, \citenamefont {Chang}, \citenamefont {Larsen}, \citenamefont
		{Fujita}, \citenamefont {Oda}, \citenamefont {Ido}, \citenamefont {Momono},
		\citenamefont {Forgan}, \citenamefont {Holmes}, \citenamefont {Mesot},
		\citenamefont {Huecker},\ and\ \citenamefont
		{v.~Zimmermann}}]{Christensen14}%
	\BibitemOpen
	\bibfield  {author} {\bibinfo {author} {\bibfnamefont {N.~B.}\ \bibnamefont
			{Christensen}}, \bibinfo {author} {\bibfnamefont {J.}~\bibnamefont {Chang}},
		\bibinfo {author} {\bibfnamefont {J.}~\bibnamefont {Larsen}}, \bibinfo
		{author} {\bibfnamefont {M.}~\bibnamefont {Fujita}}, \bibinfo {author}
		{\bibfnamefont {M.}~\bibnamefont {Oda}}, \bibinfo {author} {\bibfnamefont
			{M.}~\bibnamefont {Ido}}, \bibinfo {author} {\bibfnamefont {N.}~\bibnamefont
			{Momono}}, \bibinfo {author} {\bibfnamefont {E.~M.}\ \bibnamefont {Forgan}},
		\bibinfo {author} {\bibfnamefont {A.~T.}\ \bibnamefont {Holmes}}, \bibinfo
		{author} {\bibfnamefont {J.}~\bibnamefont {Mesot}}, \bibinfo {author}
		{\bibfnamefont {M.}~\bibnamefont {Huecker}}, \ and\ \bibinfo {author}
		{\bibfnamefont {M.}~\bibnamefont {v.~Zimmermann}},\ }\bibfield  {title}
	{\enquote {\bibinfo {title} {Bulk charge stripe order competing with
				superconductivity in
				{${\text{La}}_{2\ensuremath{-}x}$${\text{Sr}}_{x}$${\text{CuO}}_{4}(x=0.12)$}},}\
	}\href {https://arxiv.org/abs/1404.3192} {\bibinfo  {journal}
		{arXiv:1404.3192}\ }\BibitemShut {NoStop}%
	\bibitem [{\citenamefont {Thampy}\ \emph {et~al.}(2014)\citenamefont {Thampy},
		\citenamefont {Dean}, \citenamefont {Christensen}, \citenamefont {Steinke},
		\citenamefont {Islam}, \citenamefont {Oda}, \citenamefont {Ido},
		\citenamefont {Momono}, \citenamefont {Wilkins},\ and\ \citenamefont
		{Hill}}]{ThampyPRB2014}%
	\BibitemOpen
	\bibfield  {journal} {  }\bibfield  {author} {\bibinfo {author} {\bibfnamefont
			{V.}~\bibnamefont {Thampy}}, \bibinfo {author} {\bibfnamefont {M.~P.~M.}\
			\bibnamefont {Dean}}, \bibinfo {author} {\bibfnamefont {N.~B.}\ \bibnamefont
			{Christensen}}, \bibinfo {author} {\bibfnamefont {L.}~\bibnamefont
			{Steinke}}, \bibinfo {author} {\bibfnamefont {Z.}~\bibnamefont {Islam}},
		\bibinfo {author} {\bibfnamefont {M.}~\bibnamefont {Oda}}, \bibinfo {author}
		{\bibfnamefont {M.}~\bibnamefont {Ido}}, \bibinfo {author} {\bibfnamefont
			{N.}~\bibnamefont {Momono}}, \bibinfo {author} {\bibfnamefont {S.~B.}\
			\bibnamefont {Wilkins}}, \ and\ \bibinfo {author} {\bibfnamefont {J.~P.}\
			\bibnamefont {Hill}},\ }\bibfield  {title} {\enquote {\bibinfo {title}
			{Rotated stripe order and its competition with superconductivity in
				{La}$_{1.88}${Sr}$_{0.12}${CuO}$_4$},}\ }\href {\doibase
		10.1103/PhysRevB.90.100510} {\bibfield  {journal} {\bibinfo  {journal} {Phys.
				Rev. B}\ }\textbf {\bibinfo {volume} {90}},\ \bibinfo {pages} {100510(R)}
		(\bibinfo {year} {2014})}\BibitemShut {NoStop}%
	\bibitem [{\citenamefont {Kimura}\ \emph {et~al.}(2000)\citenamefont {Kimura},
		\citenamefont {Matsushita}, \citenamefont {Hirota}, \citenamefont {Endoh},
		\citenamefont {Yamada}, \citenamefont {Shirane}, \citenamefont {Lee},
		\citenamefont {Kastner},\ and\ \citenamefont {Birgeneau}}]{KimuraPRB2000}%
	\BibitemOpen
	\bibfield  {author} {\bibinfo {author} {\bibfnamefont {H.}~\bibnamefont
			{Kimura}}, \bibinfo {author} {\bibfnamefont {H.}~\bibnamefont {Matsushita}},
		\bibinfo {author} {\bibfnamefont {K.}~\bibnamefont {Hirota}}, \bibinfo
		{author} {\bibfnamefont {Y.}~\bibnamefont {Endoh}}, \bibinfo {author}
		{\bibfnamefont {K.}~\bibnamefont {Yamada}}, \bibinfo {author} {\bibfnamefont
			{G.}~\bibnamefont {Shirane}}, \bibinfo {author} {\bibfnamefont {Y.~S.}\
			\bibnamefont {Lee}}, \bibinfo {author} {\bibfnamefont {M.~A.}\ \bibnamefont
			{Kastner}}, \ and\ \bibinfo {author} {\bibfnamefont {R.~J.}\ \bibnamefont
			{Birgeneau}},\ }\bibfield  {title} {\enquote {\bibinfo {title}
			{Incommensurate geometry of the elastic magnetic peaks in superconducting
				{${\mathrm{La}}_{1.88}{\mathrm{Sr}}_{0.12}{\mathrm{CuO}}_{4}$}},}\ }\href
	{\doibase 10.1103/PhysRevB.61.14366} {\bibfield  {journal} {\bibinfo
			{journal} {Phys. Rev. B}\ }\textbf {\bibinfo {volume} {61}},\ \bibinfo
		{pages} {14366--14369} (\bibinfo {year} {2000})}\BibitemShut {NoStop}%
	\bibitem [{\citenamefont {Wakimoto}\ \emph {et~al.}(2006)\citenamefont
		{Wakimoto}, \citenamefont {Kimura}, \citenamefont {Fujita}, \citenamefont
		{Yamada}, \citenamefont {Noda}, \citenamefont {Shirane}, \citenamefont {Gu},
		\citenamefont {Kim},\ and\ \citenamefont {J.~Birgeneau}}]{WakimotoJPSJ2006}%
	\BibitemOpen
	\bibfield  {author} {\bibinfo {author} {\bibfnamefont {S.}~\bibnamefont
			{Wakimoto}}, \bibinfo {author} {\bibfnamefont {H.}~\bibnamefont {Kimura}},
		\bibinfo {author} {\bibfnamefont {M.}~\bibnamefont {Fujita}}, \bibinfo
		{author} {\bibfnamefont {K.}~\bibnamefont {Yamada}}, \bibinfo {author}
		{\bibfnamefont {Y.}~\bibnamefont {Noda}}, \bibinfo {author} {\bibfnamefont
			{G.}~\bibnamefont {Shirane}}, \bibinfo {author} {\bibfnamefont
			{G.}~\bibnamefont {Gu}}, \bibinfo {author} {\bibfnamefont {H.}~\bibnamefont
			{Kim}}, \ and\ \bibinfo {author} {\bibfnamefont {R.}~\bibnamefont
			{J.~Birgeneau}},\ }\bibfield  {title} {\enquote {\bibinfo {title}
			{Incommensurate lattice distortion in the high temperature tetragonal phase
				of
				{${\text{La}}_{2\ensuremath{-}x}$${\text{(Sr,Ba)}}_{x}$${\text{CuO}}_{4}$}},}\
	}\href {\doibase 10.1143/JPSJ.75.074714} {\bibfield  {journal} {\bibinfo
			{journal} {J. Phys. Soc. Jpn.}\ }\textbf {\bibinfo {volume} {75}},\ \bibinfo
		{pages} {074714} (\bibinfo {year} {2006})}\BibitemShut {NoStop}%
	\bibitem [{\citenamefont {McMillan}(1976)}]{McMillanPRB1976}%
	\BibitemOpen
	\bibfield  {author} {\bibinfo {author} {\bibfnamefont {W.~L.}\ \bibnamefont
			{McMillan}},\ }\bibfield  {title} {\enquote {\bibinfo {title} {Theory of
				discommensurations and the commensurate-incommensurate charge-density-wave
				phase transition},}\ }\href {\doibase 10.1103/PhysRevB.14.1496} {\bibfield
		{journal} {\bibinfo  {journal} {Phys. Rev. B}\ }\textbf {\bibinfo {volume}
			{14}},\ \bibinfo {pages} {1496--1502} (\bibinfo {year} {1976})}\BibitemShut
	{NoStop}%
	\bibitem [{\citenamefont {Eskes}\ \emph {et~al.}(1998)\citenamefont {Eskes},
		\citenamefont {Osman}, \citenamefont {Grimberg}, \citenamefont {van
			Saarloos},\ and\ \citenamefont {Zaanen}}]{EskesPRB1998}%
	\BibitemOpen
	\bibfield  {author} {\bibinfo {author} {\bibfnamefont {H.}~\bibnamefont
			{Eskes}}, \bibinfo {author} {\bibfnamefont {O.~Y.}\ \bibnamefont {Osman}},
		\bibinfo {author} {\bibfnamefont {R.}~\bibnamefont {Grimberg}}, \bibinfo
		{author} {\bibfnamefont {W.}~\bibnamefont {van Saarloos}}, \ and\ \bibinfo
		{author} {\bibfnamefont {J.}~\bibnamefont {Zaanen}},\ }\bibfield  {title}
	{\enquote {\bibinfo {title} {Charged domain walls as quantum strings on a
				lattice},}\ }\href {\doibase 10.1103/PhysRevB.58.6963} {\bibfield  {journal}
		{\bibinfo  {journal} {Phys. Rev. B}\ }\textbf {\bibinfo {volume} {58}},\
		\bibinfo {pages} {6963--6981} (\bibinfo {year} {1998})}\BibitemShut {NoStop}%
	\bibitem [{\citenamefont {Kim}\ \emph {et~al.}(2021)\citenamefont {Kim},
		\citenamefont {Lefran\ifmmode~\mbox{\c{c}}\else \c{c}\fi{}ois}, \citenamefont
		{Kummer}, \citenamefont {Fumagalli}, \citenamefont {Brookes}, \citenamefont
		{Betto}, \citenamefont {Nakata}, \citenamefont {Tortora}, \citenamefont
		{Porras}, \citenamefont {Loew}, \citenamefont {Barber}, \citenamefont
		{Braicovich}, \citenamefont {Mackenzie}, \citenamefont {Hicks}, \citenamefont
		{Keimer}, \citenamefont {Minola},\ and\ \citenamefont
		{Le~Tacon}}]{KimPRL2021}%
	\BibitemOpen
	\bibfield  {author} {\bibinfo {author} {\bibfnamefont {H.-H.}\ \bibnamefont
			{Kim}}, \bibinfo {author} {\bibfnamefont {E.}~\bibnamefont
			{Lefran\ifmmode~\mbox{\c{c}}\else \c{c}\fi{}ois}}, \bibinfo {author}
		{\bibfnamefont {K.}~\bibnamefont {Kummer}}, \bibinfo {author} {\bibfnamefont
			{R.}~\bibnamefont {Fumagalli}}, \bibinfo {author} {\bibfnamefont {N.~B.}\
			\bibnamefont {Brookes}}, \bibinfo {author} {\bibfnamefont {D.}~\bibnamefont
			{Betto}}, \bibinfo {author} {\bibfnamefont {S.}~\bibnamefont {Nakata}},
		\bibinfo {author} {\bibfnamefont {M.}~\bibnamefont {Tortora}}, \bibinfo
		{author} {\bibfnamefont {J.}~\bibnamefont {Porras}}, \bibinfo {author}
		{\bibfnamefont {T.}~\bibnamefont {Loew}}, \bibinfo {author} {\bibfnamefont
			{M.~E.}\ \bibnamefont {Barber}}, \bibinfo {author} {\bibfnamefont
			{L.}~\bibnamefont {Braicovich}}, \bibinfo {author} {\bibfnamefont {A.~P.}\
			\bibnamefont {Mackenzie}}, \bibinfo {author} {\bibfnamefont {C.~W.}\
			\bibnamefont {Hicks}}, \bibinfo {author} {\bibfnamefont {B.}~\bibnamefont
			{Keimer}}, \bibinfo {author} {\bibfnamefont {M.}~\bibnamefont {Minola}}, \
		and\ \bibinfo {author} {\bibfnamefont {M.}~\bibnamefont {Le~Tacon}},\
	}\bibfield  {title} {\enquote {\bibinfo {title} {Charge density waves in
				{YBa}$_2${Cu}$_3${O}$_{6.67}$ probed by resonant x-ray scattering under
				uniaxial compression},}\ }\href {\doibase 10.1103/PhysRevLett.126.037002}
	{\bibfield  {journal} {\bibinfo  {journal} {Phys. Rev. Lett.}\ }\textbf
		{\bibinfo {volume} {126}},\ \bibinfo {pages} {037002} (\bibinfo {year}
		{2021})}\BibitemShut {NoStop}%
	\bibitem [{\citenamefont {Lu}\ \emph {et~al.}()\citenamefont {Lu},
		\citenamefont {Zhang}, \citenamefont {Tseng}, \citenamefont {Liu},
		\citenamefont {Tao}, \citenamefont {Paris}, \citenamefont {Liu},
		\citenamefont {Chen}, \citenamefont {Strocov}, \citenamefont {Song},
		\citenamefont {Yu}, \citenamefont {Si}, \citenamefont {Dai},\ and\
		\citenamefont {Schmitt}}]{Lu21}%
	\BibitemOpen
	\bibfield  {author} {\bibinfo {author} {\bibfnamefont {X.}~\bibnamefont
			{Lu}}, \bibinfo {author} {\bibfnamefont {W.}~\bibnamefont {Zhang}}, \bibinfo
		{author} {\bibfnamefont {Y.}~\bibnamefont {Tseng}}, \bibinfo {author}
		{\bibfnamefont {R.}~\bibnamefont {Liu}}, \bibinfo {author} {\bibfnamefont
			{Z.}~\bibnamefont {Tao}}, \bibinfo {author} {\bibfnamefont {E.}~\bibnamefont
			{Paris}}, \bibinfo {author} {\bibfnamefont {P.}~\bibnamefont {Liu}}, \bibinfo
		{author} {\bibfnamefont {T.}~\bibnamefont {Chen}}, \bibinfo {author}
		{\bibfnamefont {V.}~\bibnamefont {Strocov}}, \bibinfo {author} {\bibfnamefont
			{Y.}~\bibnamefont {Song}}, \bibinfo {author} {\bibfnamefont {R.}~\bibnamefont
			{Yu}}, \bibinfo {author} {\bibfnamefont {Q.}~\bibnamefont {Si}}, \bibinfo
		{author} {\bibfnamefont {P.}~\bibnamefont {Dai}}, \ and\ \bibinfo {author}
		{\bibfnamefont {T.}~\bibnamefont {Schmitt}},\ }\bibfield  {title} {\enquote
		{\bibinfo {title} {{Spin-excitation anisotropy in the nematic state of
					detwinned {FeSe}}},}\ }\href {http://arxiv.org/abs/2108.04484} {\bibinfo
		{journal} {arXiv:2107.10264}\ }\BibitemShut {NoStop}%
	\bibitem [{\citenamefont {Ivashko}\ \emph {et~al.}(2019)\citenamefont
		{Ivashko}, \citenamefont {Horio}, \citenamefont {Wan}, \citenamefont
		{Christensen}, \citenamefont {McNally}, \citenamefont {Paris}, \citenamefont
		{Tseng}, \citenamefont {Shaik}, \citenamefont {R{\o}nnow}, \citenamefont
		{Wei}, \citenamefont {Adamo}, \citenamefont {Lichtensteiger}, \citenamefont
		{Gibert}, \citenamefont {Beasley}, \citenamefont {Shen}, \citenamefont
		{Tomczak}, \citenamefont {Schmitt},\ and\ \citenamefont
		{Chang}}]{IvashkoNC2019}%
	\BibitemOpen
	\bibfield  {journal} {  }\bibfield  {author} {\bibinfo {author} {\bibfnamefont
			{O.}~\bibnamefont {Ivashko}}, \bibinfo {author} {\bibfnamefont
			{M.}~\bibnamefont {Horio}}, \bibinfo {author} {\bibfnamefont
			{W.}~\bibnamefont {Wan}}, \bibinfo {author} {\bibfnamefont {N.~B.}\
			\bibnamefont {Christensen}}, \bibinfo {author} {\bibfnamefont {D.~E.}\
			\bibnamefont {McNally}}, \bibinfo {author} {\bibfnamefont {E.}~\bibnamefont
			{Paris}}, \bibinfo {author} {\bibfnamefont {Y.}~\bibnamefont {Tseng}},
		\bibinfo {author} {\bibfnamefont {N.~E.}\ \bibnamefont {Shaik}}, \bibinfo
		{author} {\bibfnamefont {H.~M.}\ \bibnamefont {R{\o}nnow}}, \bibinfo {author}
		{\bibfnamefont {H.~I.}\ \bibnamefont {Wei}}, \bibinfo {author} {\bibfnamefont
			{C.}~\bibnamefont {Adamo}}, \bibinfo {author} {\bibfnamefont
			{C.}~\bibnamefont {Lichtensteiger}}, \bibinfo {author} {\bibfnamefont
			{M.}~\bibnamefont {Gibert}}, \bibinfo {author} {\bibfnamefont {M.~R.}\
			\bibnamefont {Beasley}}, \bibinfo {author} {\bibfnamefont {K.~M.}\
			\bibnamefont {Shen}}, \bibinfo {author} {\bibfnamefont {J.~M.}\ \bibnamefont
			{Tomczak}}, \bibinfo {author} {\bibfnamefont {T.}~\bibnamefont {Schmitt}}, \
		and\ \bibinfo {author} {\bibfnamefont {J.}~\bibnamefont {Chang}},\ }\bibfield
	{title} {\enquote {\bibinfo {title} {{Strain-engineering Mott-insulating
					La$_2$CuO$_4$}},}\ }\href {\doibase 10.1038/s41467-019-08664-6} {\bibfield
		{journal} {\bibinfo  {journal} {Nat. Commun.}\ }\textbf {\bibinfo {volume}
			{10}},\ \bibinfo {pages} {786} (\bibinfo {year} {2019})}\BibitemShut
	{NoStop}%
	\bibitem [{\citenamefont {Wang}\ \emph {et~al.}(2019)\citenamefont {Wang},
		\citenamefont {Geessinck}, \citenamefont {Elnaggar}, \citenamefont
		{Birkh\"olzer}, \citenamefont {Tomiyasu}, \citenamefont {Okamoto},
		\citenamefont {Liu}, \citenamefont {Du}, \citenamefont {Huang}, \citenamefont
		{Koster},\ and\ \citenamefont {de~Groot}}]{WangPRB2019}%
	\BibitemOpen
	\bibfield  {author} {\bibinfo {author} {\bibfnamefont {R.-P.}\ \bibnamefont
			{Wang}}, \bibinfo {author} {\bibfnamefont {J.}~\bibnamefont {Geessinck}},
		\bibinfo {author} {\bibfnamefont {H.}~\bibnamefont {Elnaggar}}, \bibinfo
		{author} {\bibfnamefont {Y.~A.}\ \bibnamefont {Birkh\"olzer}}, \bibinfo
		{author} {\bibfnamefont {K.}~\bibnamefont {Tomiyasu}}, \bibinfo {author}
		{\bibfnamefont {J.}~\bibnamefont {Okamoto}}, \bibinfo {author} {\bibfnamefont
			{B.}~\bibnamefont {Liu}}, \bibinfo {author} {\bibfnamefont {C.-H.}\
			\bibnamefont {Du}}, \bibinfo {author} {\bibfnamefont {D.-J.}\ \bibnamefont
			{Huang}}, \bibinfo {author} {\bibfnamefont {G.}~\bibnamefont {Koster}}, \
		and\ \bibinfo {author} {\bibfnamefont {F.~M.~F.}\ \bibnamefont {de~Groot}},\
	}\bibfield  {title} {\enquote {\bibinfo {title} {Low-energy orbital
				excitations in strained {LaCoO}$_{3}$ films},}\ }\href {\doibase
		10.1103/PhysRevB.100.165148} {\bibfield  {journal} {\bibinfo  {journal}
			{Phys. Rev. B}\ }\textbf {\bibinfo {volume} {100}},\ \bibinfo {pages}
		{165148} (\bibinfo {year} {2019})}\BibitemShut {NoStop}%
	\bibitem [{\citenamefont {Paris}\ \emph {et~al.}(2020)\citenamefont {Paris},
		\citenamefont {Tseng}, \citenamefont {P\"{a}rschke}, \citenamefont {Zhang},
		\citenamefont {Upton}, \citenamefont {Efimenko}, \citenamefont {Rolfs},
		\citenamefont {McNally}, \citenamefont {Maurel}, \citenamefont {Naamneh},
		\citenamefont {Caputo}, \citenamefont {Strocov}, \citenamefont {Wang},
		\citenamefont {Casa}, \citenamefont {Schneider}, \citenamefont
		{Pomjakushina}, \citenamefont {Wohlfeld}, \citenamefont {Radovic},\ and\
		\citenamefont {Schmitt}}]{ParisPNAS2020}%
	\BibitemOpen
	\bibfield  {author} {\bibinfo {author} {\bibfnamefont {E.}~\bibnamefont
			{Paris}}, \bibinfo {author} {\bibfnamefont {Y.}~\bibnamefont {Tseng}},
		\bibinfo {author} {\bibfnamefont {E.~M.}\ \bibnamefont {P\"{a}rschke}},
		\bibinfo {author} {\bibfnamefont {W.}~\bibnamefont {Zhang}}, \bibinfo
		{author} {\bibfnamefont {M.~H.}\ \bibnamefont {Upton}}, \bibinfo {author}
		{\bibfnamefont {A.}~\bibnamefont {Efimenko}}, \bibinfo {author}
		{\bibfnamefont {K.}~\bibnamefont {Rolfs}}, \bibinfo {author} {\bibfnamefont
			{D.~E.}\ \bibnamefont {McNally}}, \bibinfo {author} {\bibfnamefont
			{L.}~\bibnamefont {Maurel}}, \bibinfo {author} {\bibfnamefont
			{M.}~\bibnamefont {Naamneh}}, \bibinfo {author} {\bibfnamefont
			{M.}~\bibnamefont {Caputo}}, \bibinfo {author} {\bibfnamefont {V.~N.}\
			\bibnamefont {Strocov}}, \bibinfo {author} {\bibfnamefont {Z.}~\bibnamefont
			{Wang}}, \bibinfo {author} {\bibfnamefont {D.}~\bibnamefont {Casa}}, \bibinfo
		{author} {\bibfnamefont {C.~W.}\ \bibnamefont {Schneider}}, \bibinfo {author}
		{\bibfnamefont {E.}~\bibnamefont {Pomjakushina}}, \bibinfo {author}
		{\bibfnamefont {K.}~\bibnamefont {Wohlfeld}}, \bibinfo {author}
		{\bibfnamefont {M.}~\bibnamefont {Radovic}}, \ and\ \bibinfo {author}
		{\bibfnamefont {T.}~\bibnamefont {Schmitt}},\ }\bibfield  {title} {\enquote
		{\bibinfo {title} {Strain engineering of the charge and spin-orbital
				interactions in {Sr}$_2${IrO}$_4$},}\ }\href {\doibase
		10.1073/pnas.2012043117} {\bibfield  {journal} {\bibinfo  {journal} {Proc.
				Natl. Acad. Sci. U.S.A.}\ }\textbf {\bibinfo {volume} {117}},\ \bibinfo
		{pages} {24764--24770} (\bibinfo {year} {2020})}\BibitemShut {NoStop}%
	\bibitem [{\citenamefont {Moretti~Sala}\ \emph {et~al.}(2011)\citenamefont
		{Moretti~Sala}, \citenamefont {Bisogni}, \citenamefont {Aruta}, \citenamefont
		{Balestrino}, \citenamefont {Berger}, \citenamefont {Brookes}, \citenamefont
		{Luca}, \citenamefont {Di~Castro}, \citenamefont {Grioni}, \citenamefont
		{Guarise}, \citenamefont {Medaglia}, \citenamefont {Miletto~Granozio},
		\citenamefont {Minola}, \citenamefont {Perna}, \citenamefont {Radovic},
		\citenamefont {Salluzzo}, \citenamefont {Schmitt}, \citenamefont {Zhou},
		\citenamefont {Braicovich},\ and\ \citenamefont
		{Ghiringhelli}}]{SalaNJP2011}%
	\BibitemOpen
	\bibfield  {author} {\bibinfo {author} {\bibfnamefont {M.}~\bibnamefont
			{Moretti~Sala}}, \bibinfo {author} {\bibfnamefont {V.}~\bibnamefont
			{Bisogni}}, \bibinfo {author} {\bibfnamefont {C.}~\bibnamefont {Aruta}},
		\bibinfo {author} {\bibfnamefont {G.}~\bibnamefont {Balestrino}}, \bibinfo
		{author} {\bibfnamefont {H.}~\bibnamefont {Berger}}, \bibinfo {author}
		{\bibfnamefont {N.~B.}\ \bibnamefont {Brookes}}, \bibinfo {author}
		{\bibfnamefont {G.~M.~d.}\ \bibnamefont {Luca}}, \bibinfo {author}
		{\bibfnamefont {D.}~\bibnamefont {Di~Castro}}, \bibinfo {author}
		{\bibfnamefont {M.}~\bibnamefont {Grioni}}, \bibinfo {author} {\bibfnamefont
			{M.}~\bibnamefont {Guarise}}, \bibinfo {author} {\bibfnamefont {P.~G.}\
			\bibnamefont {Medaglia}}, \bibinfo {author} {\bibfnamefont {F.}~\bibnamefont
			{Miletto~Granozio}}, \bibinfo {author} {\bibfnamefont {M.}~\bibnamefont
			{Minola}}, \bibinfo {author} {\bibfnamefont {P.}~\bibnamefont {Perna}},
		\bibinfo {author} {\bibfnamefont {M.}~\bibnamefont {Radovic}}, \bibinfo
		{author} {\bibfnamefont {M.}~\bibnamefont {Salluzzo}}, \bibinfo {author}
		{\bibfnamefont {T.}~\bibnamefont {Schmitt}}, \bibinfo {author} {\bibfnamefont
			{K.~J.}\ \bibnamefont {Zhou}}, \bibinfo {author} {\bibfnamefont
			{L.}~\bibnamefont {Braicovich}}, \ and\ \bibinfo {author} {\bibfnamefont
			{G.}~\bibnamefont {Ghiringhelli}},\ }\bibfield  {title} {\enquote {\bibinfo
			{title} {Energy and symmetry of dd excitations in undoped layered cuprates
				measured by {Cu} \textit{{L}}$_{\textrm{3}}$ resonant inelastic x-ray
				scattering},}\ }\href {\doibase 10.1088/1367-2630/13/4/043026} {\bibfield
		{journal} {\bibinfo  {journal} {New J. Phys.}\ }\textbf {\bibinfo {volume}
			{13}},\ \bibinfo {pages} {043026} (\bibinfo {year} {2011})}\BibitemShut
	{NoStop}%
	\bibitem [{\citenamefont {Choi}\ \emph {et~al.}()\citenamefont {Choi},
		\citenamefont {Wang}, \citenamefont {Jöhr}, \citenamefont {Christensen},
		\citenamefont {Küspert}, \citenamefont {Bucher}, \citenamefont {Biscette},
		\citenamefont {Hücker}, \citenamefont {Kurosawa}, \citenamefont {Momono},
		\citenamefont {Oda}, \citenamefont {Ivashko}, \citenamefont {v.~Zimmermann},
		\citenamefont {Janoschek},\ and\ \citenamefont {Chang}}]{Choi20}%
	\BibitemOpen
	\bibfield  {author} {\bibinfo {author} {\bibfnamefont {J.}~\bibnamefont
			{Choi}}, \bibinfo {author} {\bibfnamefont {Q.}~\bibnamefont {Wang}}, \bibinfo
		{author} {\bibfnamefont {S.}~\bibnamefont {Jöhr}}, \bibinfo {author}
		{\bibfnamefont {N.~B.}\ \bibnamefont {Christensen}}, \bibinfo {author}
		{\bibfnamefont {J.}~\bibnamefont {Küspert}}, \bibinfo {author}
		{\bibfnamefont {D.}~\bibnamefont {Bucher}}, \bibinfo {author} {\bibfnamefont
			{D.}~\bibnamefont {Biscette}}, \bibinfo {author} {\bibfnamefont
			{M.}~\bibnamefont {Hücker}}, \bibinfo {author} {\bibfnamefont
			{T.}~\bibnamefont {Kurosawa}}, \bibinfo {author} {\bibfnamefont
			{N.}~\bibnamefont {Momono}}, \bibinfo {author} {\bibfnamefont
			{M.}~\bibnamefont {Oda}}, \bibinfo {author} {\bibfnamefont {O.}~\bibnamefont
			{Ivashko}}, \bibinfo {author} {\bibfnamefont {M.}~\bibnamefont
			{v.~Zimmermann}}, \bibinfo {author} {\bibfnamefont {M.}~\bibnamefont
			{Janoschek}}, \ and\ \bibinfo {author} {\bibfnamefont {J.}~\bibnamefont
			{Chang}},\ }\bibfield  {title} {\enquote {\bibinfo {title} {Disentangling
				intertwined quantum states in a prototypical cuprate superconductor},}\
	}\href {https://arxiv.org/abs/2009.06967} {\bibinfo  {journal}
		{arXiv:2009.06967}\ }\BibitemShut {NoStop}%
	\bibitem [{\citenamefont {Meyer}\ \emph {et~al.}(2015)\citenamefont {Meyer},
		\citenamefont {Jiang}, \citenamefont {Park}, \citenamefont {Egami},\ and\
		\citenamefont {Lee}}]{MeyerAPL2015}%
	\BibitemOpen
	\bibfield  {journal} {  }\bibfield  {author} {\bibinfo {author} {\bibfnamefont
			{T.~L.}\ \bibnamefont {Meyer}}, \bibinfo {author} {\bibfnamefont
			{L.}~\bibnamefont {Jiang}}, \bibinfo {author} {\bibfnamefont
			{S.}~\bibnamefont {Park}}, \bibinfo {author} {\bibfnamefont {T.}~\bibnamefont
			{Egami}}, \ and\ \bibinfo {author} {\bibfnamefont {H.~N.}\ \bibnamefont
			{Lee}},\ }\bibfield  {title} {\enquote {\bibinfo {title} {Strain-relaxation
				and critical thickness of epitaxial {La}$_{1.85}${Sr}$_{0.15}${CuO}$_4$
				films},}\ }\href {\doibase 10.1063/1.4937170} {\bibfield  {journal} {\bibinfo
			{journal} {{APL} Materials}\ }\textbf {\bibinfo {volume} {3}},\ \bibinfo
		{pages} {126102} (\bibinfo {year} {2015})}\BibitemShut {NoStop}%
	\bibitem [{\citenamefont {Wang}\ \emph {et~al.}(2020)\citenamefont {Wang},
		\citenamefont {Horio}, \citenamefont {von Arx}, \citenamefont {Shen},
		\citenamefont {John~Mukkattukavil}, \citenamefont {Sassa}, \citenamefont
		{Ivashko}, \citenamefont {Matt}, \citenamefont {Pyon}, \citenamefont
		{Takayama}, \citenamefont {Takagi}, \citenamefont {Kurosawa}, \citenamefont
		{Momono}, \citenamefont {Oda}, \citenamefont {Adachi}, \citenamefont
		{Haidar}, \citenamefont {Koike}, \citenamefont {Tseng}, \citenamefont
		{Zhang}, \citenamefont {Zhao}, \citenamefont {Kummer}, \citenamefont
		{Garcia-Fernandez}, \citenamefont {Zhou}, \citenamefont {Christensen},
		\citenamefont {R\o{}nnow}, \citenamefont {Schmitt},\ and\ \citenamefont
		{Chang}}]{WangPRL2020}%
	\BibitemOpen
	\bibfield  {author} {\bibinfo {author} {\bibfnamefont {Q.}~\bibnamefont
			{Wang}}, \bibinfo {author} {\bibfnamefont {M.}~\bibnamefont {Horio}},
		\bibinfo {author} {\bibfnamefont {K.}~\bibnamefont {von Arx}}, \bibinfo
		{author} {\bibfnamefont {Y.}~\bibnamefont {Shen}}, \bibinfo {author}
		{\bibfnamefont {D.}~\bibnamefont {John~Mukkattukavil}}, \bibinfo {author}
		{\bibfnamefont {Y.}~\bibnamefont {Sassa}}, \bibinfo {author} {\bibfnamefont
			{O.}~\bibnamefont {Ivashko}}, \bibinfo {author} {\bibfnamefont {C.~E.}\
			\bibnamefont {Matt}}, \bibinfo {author} {\bibfnamefont {S.}~\bibnamefont
			{Pyon}}, \bibinfo {author} {\bibfnamefont {T.}~\bibnamefont {Takayama}},
		\bibinfo {author} {\bibfnamefont {H.}~\bibnamefont {Takagi}}, \bibinfo
		{author} {\bibfnamefont {T.}~\bibnamefont {Kurosawa}}, \bibinfo {author}
		{\bibfnamefont {N.}~\bibnamefont {Momono}}, \bibinfo {author} {\bibfnamefont
			{M.}~\bibnamefont {Oda}}, \bibinfo {author} {\bibfnamefont {T.}~\bibnamefont
			{Adachi}}, \bibinfo {author} {\bibfnamefont {S.~M.}\ \bibnamefont {Haidar}},
		\bibinfo {author} {\bibfnamefont {Y.}~\bibnamefont {Koike}}, \bibinfo
		{author} {\bibfnamefont {Y.}~\bibnamefont {Tseng}}, \bibinfo {author}
		{\bibfnamefont {W.}~\bibnamefont {Zhang}}, \bibinfo {author} {\bibfnamefont
			{J.}~\bibnamefont {Zhao}}, \bibinfo {author} {\bibfnamefont {K.}~\bibnamefont
			{Kummer}}, \bibinfo {author} {\bibfnamefont {M.}~\bibnamefont
			{Garcia-Fernandez}}, \bibinfo {author} {\bibfnamefont {K.-J.}\ \bibnamefont
			{Zhou}}, \bibinfo {author} {\bibfnamefont {N.~B.}\ \bibnamefont
			{Christensen}}, \bibinfo {author} {\bibfnamefont {H.~M.}\ \bibnamefont
			{R\o{}nnow}}, \bibinfo {author} {\bibfnamefont {T.}~\bibnamefont {Schmitt}},
		\ and\ \bibinfo {author} {\bibfnamefont {J.}~\bibnamefont {Chang}},\
	}\bibfield  {title} {\enquote {\bibinfo {title} {High-temperature
				charge-stripe correlations in
				{${\mathrm{La}}_{1.675}{\mathrm{Eu}}_{0.2}{\mathrm{Sr}}_{0.125}{\mathrm{CuO}}_{4}$}},}\
	}\href {\doibase 10.1103/PhysRevLett.124.187002} {\bibfield  {journal}
		{\bibinfo  {journal} {Phys. Rev. Lett.}\ }\textbf {\bibinfo {volume} {124}},\
		\bibinfo {pages} {187002} (\bibinfo {year} {2020})}\BibitemShut {NoStop}%
	\bibitem [{\citenamefont {Ivashko}\ \emph {et~al.}(2017)\citenamefont
		{Ivashko}, \citenamefont {Shaik}, \citenamefont {Lu}, \citenamefont
		{Fatuzzo}, \citenamefont {Dantz}, \citenamefont {Freeman}, \citenamefont
		{McNally}, \citenamefont {Destraz}, \citenamefont {Christensen},
		\citenamefont {Kurosawa}, \citenamefont {Momono}, \citenamefont {Oda},
		\citenamefont {Matt}, \citenamefont {Monney}, \citenamefont {R\o{}nnow},
		\citenamefont {Schmitt},\ and\ \citenamefont {Chang}}]{IvashkoPRB2017}%
	\BibitemOpen
	\bibfield  {author} {\bibinfo {author} {\bibfnamefont {O.}~\bibnamefont
			{Ivashko}}, \bibinfo {author} {\bibfnamefont {N.~E.}\ \bibnamefont {Shaik}},
		\bibinfo {author} {\bibfnamefont {X.}~\bibnamefont {Lu}}, \bibinfo {author}
		{\bibfnamefont {C.~G.}\ \bibnamefont {Fatuzzo}}, \bibinfo {author}
		{\bibfnamefont {M.}~\bibnamefont {Dantz}}, \bibinfo {author} {\bibfnamefont
			{P.~G.}\ \bibnamefont {Freeman}}, \bibinfo {author} {\bibfnamefont {D.~E.}\
			\bibnamefont {McNally}}, \bibinfo {author} {\bibfnamefont {D.}~\bibnamefont
			{Destraz}}, \bibinfo {author} {\bibfnamefont {N.~B.}\ \bibnamefont
			{Christensen}}, \bibinfo {author} {\bibfnamefont {T.}~\bibnamefont
			{Kurosawa}}, \bibinfo {author} {\bibfnamefont {N.}~\bibnamefont {Momono}},
		\bibinfo {author} {\bibfnamefont {M.}~\bibnamefont {Oda}}, \bibinfo {author}
		{\bibfnamefont {C.~E.}\ \bibnamefont {Matt}}, \bibinfo {author}
		{\bibfnamefont {C.}~\bibnamefont {Monney}}, \bibinfo {author} {\bibfnamefont
			{H.~M.}\ \bibnamefont {R\o{}nnow}}, \bibinfo {author} {\bibfnamefont
			{T.}~\bibnamefont {Schmitt}}, \ and\ \bibinfo {author} {\bibfnamefont
			{J.}~\bibnamefont {Chang}},\ }\bibfield  {title} {\enquote {\bibinfo {title}
			{Damped spin excitations in a doped cuprate superconductor with orbital
				hybridization},}\ }\href {\doibase 10.1103/PhysRevB.95.214508} {\bibfield
		{journal} {\bibinfo  {journal} {Phys. Rev. B}\ }\textbf {\bibinfo {volume}
			{95}},\ \bibinfo {pages} {214508} (\bibinfo {year} {2017})}\BibitemShut
	{NoStop}%
	\bibitem [{\citenamefont {Wen}\ \emph {et~al.}(2019)\citenamefont {Wen},
		\citenamefont {Huang}, \citenamefont {Lee}, \citenamefont {Jang},
		\citenamefont {Knight}, \citenamefont {Lee}, \citenamefont {Fujita},
		\citenamefont {Suzuki}, \citenamefont {Asano}, \citenamefont {Kivelson},
		\citenamefont {Kao},\ and\ \citenamefont {Lee}}]{WenNC2019}%
	\BibitemOpen
	\bibfield  {author} {\bibinfo {author} {\bibfnamefont {J.~J.}\ \bibnamefont
			{Wen}}, \bibinfo {author} {\bibfnamefont {H.}~\bibnamefont {Huang}}, \bibinfo
		{author} {\bibfnamefont {S.~J.}\ \bibnamefont {Lee}}, \bibinfo {author}
		{\bibfnamefont {H.}~\bibnamefont {Jang}}, \bibinfo {author} {\bibfnamefont
			{J.}~\bibnamefont {Knight}}, \bibinfo {author} {\bibfnamefont {Y.~S.}\
			\bibnamefont {Lee}}, \bibinfo {author} {\bibfnamefont {M.}~\bibnamefont
			{Fujita}}, \bibinfo {author} {\bibfnamefont {K.~M.}\ \bibnamefont {Suzuki}},
		\bibinfo {author} {\bibfnamefont {S.}~\bibnamefont {Asano}}, \bibinfo
		{author} {\bibfnamefont {S.~A.}\ \bibnamefont {Kivelson}}, \bibinfo {author}
		{\bibfnamefont {C.~C.}\ \bibnamefont {Kao}}, \ and\ \bibinfo {author}
		{\bibfnamefont {J.~S.}\ \bibnamefont {Lee}},\ }\bibfield  {title} {\enquote
		{\bibinfo {title} {Observation of two types of charge-density-wave orders in
				superconducting
				{${\text{La}}_{2\ensuremath{-}x}{\text{Sr}}_{x}{\text{CuO}}_{4}$}},}\ }\href
	{\doibase 10.1038/s41467-019-11167-z} {\bibfield  {journal} {\bibinfo
			{journal} {Nat. Commun.}\ }\textbf {\bibinfo {volume} {10}},\ \bibinfo
		{pages} {3269} (\bibinfo {year} {2019})}\BibitemShut {NoStop}%
	\bibitem [{\citenamefont {Robertson}\ \emph {et~al.}(2006)\citenamefont
		{Robertson}, \citenamefont {Kivelson}, \citenamefont {Fradkin}, \citenamefont
		{Fang},\ and\ \citenamefont {Kapitulnik}}]{RobertsonPRB2006}%
	\BibitemOpen
	\bibfield  {author} {\bibinfo {author} {\bibfnamefont {J.~A.}\ \bibnamefont
			{Robertson}}, \bibinfo {author} {\bibfnamefont {S.~A.}\ \bibnamefont
			{Kivelson}}, \bibinfo {author} {\bibfnamefont {E.}~\bibnamefont {Fradkin}},
		\bibinfo {author} {\bibfnamefont {A.~C.}\ \bibnamefont {Fang}}, \ and\
		\bibinfo {author} {\bibfnamefont {A.}~\bibnamefont {Kapitulnik}},\ }\bibfield
	{title} {\enquote {\bibinfo {title} {Distinguishing patterns of charge
				order: Stripes or checkerboards},}\ }\href {\doibase
		10.1103/PhysRevB.74.134507} {\bibfield  {journal} {\bibinfo  {journal} {Phys.
				Rev. B}\ }\textbf {\bibinfo {volume} {74}},\ \bibinfo {pages} {134507}
		(\bibinfo {year} {2006})}\BibitemShut {NoStop}%
	\bibitem [{\citenamefont {Del~Maestro}\ \emph {et~al.}(2006)\citenamefont
		{Del~Maestro}, \citenamefont {Rosenow},\ and\ \citenamefont
		{Sachdev}}]{MaestroPRB2006}%
	\BibitemOpen
	\bibfield  {author} {\bibinfo {author} {\bibfnamefont {A.}~\bibnamefont
			{Del~Maestro}}, \bibinfo {author} {\bibfnamefont {B.}~\bibnamefont
			{Rosenow}}, \ and\ \bibinfo {author} {\bibfnamefont {S.}~\bibnamefont
			{Sachdev}},\ }\bibfield  {title} {\enquote {\bibinfo {title} {From stripe to
				checkerboard ordering of charge-density waves on the square lattice in the
				presence of quenched disorder},}\ }\href {\doibase
		10.1103/PhysRevB.74.024520} {\bibfield  {journal} {\bibinfo  {journal} {Phys.
				Rev. B}\ }\textbf {\bibinfo {volume} {74}},\ \bibinfo {pages} {024520}
		(\bibinfo {year} {2006})}\BibitemShut {NoStop}%
	\bibitem [{\citenamefont {{He}}\ \emph {et~al.}()\citenamefont {{He}},
		\citenamefont {{Wen}}, \citenamefont {{Jiang}}, \citenamefont {{Xu}},
		\citenamefont {{Tian}}, \citenamefont {{Taniguchi}}, \citenamefont {{Ikeda}},
		\citenamefont {{Fujita}},\ and\ \citenamefont {{Lee}}}]{He21}%
	\BibitemOpen
	\bibfield  {author} {\bibinfo {author} {\bibfnamefont {W.}~\bibnamefont
			{{He}}}, \bibinfo {author} {\bibfnamefont {J.}~\bibnamefont {{Wen}}},
		\bibinfo {author} {\bibfnamefont {H.-C.}\ \bibnamefont {{Jiang}}}, \bibinfo
		{author} {\bibfnamefont {G.}~\bibnamefont {{Xu}}}, \bibinfo {author}
		{\bibfnamefont {W.}~\bibnamefont {{Tian}}}, \bibinfo {author} {\bibfnamefont
			{T.}~\bibnamefont {{Taniguchi}}}, \bibinfo {author} {\bibfnamefont
			{Y.}~\bibnamefont {{Ikeda}}}, \bibinfo {author} {\bibfnamefont
			{M.}~\bibnamefont {{Fujita}}}, \ and\ \bibinfo {author} {\bibfnamefont
			{Y.~S.}\ \bibnamefont {{Lee}}},\ }\bibfield  {title} {\enquote {\bibinfo
			{title} {{Prevalence of tilted stripes in
					${\mathrm{La}}_{1.88}{\mathrm{Sr}}_{0.12}{\mathrm{CuO}}_{4}$ and the
					importance of $t^{\prime}$ in the Hamiltonian}},}\ }\href
	{https://arxiv.org/abs/2107.10264} {\bibinfo  {journal} {arXiv:2107.10264}\
	}\BibitemShut {NoStop}%
	\bibitem [{\citenamefont {Wang}\ \emph {et~al.}(2021)\citenamefont {Wang},
		\citenamefont {von Arx}, \citenamefont {Horio}, \citenamefont
		{Mukkattukavil}, \citenamefont {K{\"u}spert}, \citenamefont {Sassa},
		\citenamefont {Schmitt}, \citenamefont {Nag}, \citenamefont {Pyon},
		\citenamefont {Takayama}, \citenamefont {Takagi}, \citenamefont
		{Garcia-Fernandez}, \citenamefont {Zhou},\ and\ \citenamefont
		{Chang}}]{WangSciAdv2021}%
	\BibitemOpen
	\bibfield  {journal} {  }\bibfield  {author} {\bibinfo {author} {\bibfnamefont
			{Q.}~\bibnamefont {Wang}}, \bibinfo {author} {\bibfnamefont {K.}~\bibnamefont
			{von Arx}}, \bibinfo {author} {\bibfnamefont {M.}~\bibnamefont {Horio}},
		\bibinfo {author} {\bibfnamefont {D.~J.}\ \bibnamefont {Mukkattukavil}},
		\bibinfo {author} {\bibfnamefont {J.}~\bibnamefont {K{\"u}spert}}, \bibinfo
		{author} {\bibfnamefont {Y.}~\bibnamefont {Sassa}}, \bibinfo {author}
		{\bibfnamefont {T.}~\bibnamefont {Schmitt}}, \bibinfo {author} {\bibfnamefont
			{A.}~\bibnamefont {Nag}}, \bibinfo {author} {\bibfnamefont {S.}~\bibnamefont
			{Pyon}}, \bibinfo {author} {\bibfnamefont {T.}~\bibnamefont {Takayama}},
		\bibinfo {author} {\bibfnamefont {H.}~\bibnamefont {Takagi}}, \bibinfo
		{author} {\bibfnamefont {M.}~\bibnamefont {Garcia-Fernandez}}, \bibinfo
		{author} {\bibfnamefont {K.-J.}\ \bibnamefont {Zhou}}, \ and\ \bibinfo
		{author} {\bibfnamefont {J.}~\bibnamefont {Chang}},\ }\bibfield  {title}
	{\enquote {\bibinfo {title} {Charge order lock-in by electron-phonon coupling
				in
				{${\mathrm{La}}_{1.675}{\mathrm{Eu}}_{0.2}{\mathrm{Sr}}_{0.125}{\mathrm{CuO}}_{4}$}},}\
	}\href {\doibase 10.1126/sciadv.abg7394} {\bibfield  {journal} {\bibinfo
			{journal} {Sci. Adv.}\ }\textbf {\bibinfo {volume} {7}},\ \bibinfo {pages}
		{eabg7394} (\bibinfo {year} {2021})}\BibitemShut {NoStop}%
	\bibitem [{\citenamefont {Ghiringhelli}\ \emph {et~al.}(2006)\citenamefont
		{Ghiringhelli}, \citenamefont {Piazzalunga}, \citenamefont {Dallera},
		\citenamefont {Trezzi}, \citenamefont {Braicovich}, \citenamefont {Schmitt},
		\citenamefont {Strocov}, \citenamefont {Betemps}, \citenamefont {Patthey},
		\citenamefont {Wang},\ and\ \citenamefont
		{Grioni}}]{GhiringhelliREVSCIINS2006}%
	\BibitemOpen
	\bibfield  {author} {\bibinfo {author} {\bibfnamefont {G.}~\bibnamefont
			{Ghiringhelli}}, \bibinfo {author} {\bibfnamefont {A.}~\bibnamefont
			{Piazzalunga}}, \bibinfo {author} {\bibfnamefont {C.}~\bibnamefont
			{Dallera}}, \bibinfo {author} {\bibfnamefont {G.}~\bibnamefont {Trezzi}},
		\bibinfo {author} {\bibfnamefont {L.}~\bibnamefont {Braicovich}}, \bibinfo
		{author} {\bibfnamefont {T.}~\bibnamefont {Schmitt}}, \bibinfo {author}
		{\bibfnamefont {V.~N.}\ \bibnamefont {Strocov}}, \bibinfo {author}
		{\bibfnamefont {R.}~\bibnamefont {Betemps}}, \bibinfo {author} {\bibfnamefont
			{L.}~\bibnamefont {Patthey}}, \bibinfo {author} {\bibfnamefont
			{X.}~\bibnamefont {Wang}}, \ and\ \bibinfo {author} {\bibfnamefont
			{M.}~\bibnamefont {Grioni}},\ }\bibfield  {title} {\enquote {\bibinfo {title}
			{{SAXES, a high resolution spectrometer for resonant x-ray emission in the
					400-1600 eV energy range}},}\ }\href {\doibase 10.1063/1.2372731} {\bibfield
		{journal} {\bibinfo  {journal} {Rev. Sci. Instrum.}\ }\textbf {\bibinfo
			{volume} {77}},\ \bibinfo {pages} {113108} (\bibinfo {year}
		{2006})}\BibitemShut {NoStop}%
	\bibitem [{\citenamefont {Strocov}\ \emph {et~al.}(2010)\citenamefont
		{Strocov}, \citenamefont {Schmitt}, \citenamefont {Flechsig}, \citenamefont
		{Schmidt}, \citenamefont {Imhof}, \citenamefont {Chen}, \citenamefont
		{Raabe}, \citenamefont {Betemps}, \citenamefont {Zimoch}, \citenamefont
		{Krempasky}, \citenamefont {Wang}, \citenamefont {Grioni},\ and\
		\citenamefont {Patthey}}]{StrocovJSYNRAD2010}%
	\BibitemOpen
	\bibfield  {author} {\bibinfo {author} {\bibfnamefont {V.~N.}\ \bibnamefont
			{Strocov}}, \bibinfo {author} {\bibfnamefont {T.}~\bibnamefont {Schmitt}},
		\bibinfo {author} {\bibfnamefont {U.}~\bibnamefont {Flechsig}}, \bibinfo
		{author} {\bibfnamefont {T.}~\bibnamefont {Schmidt}}, \bibinfo {author}
		{\bibfnamefont {A.}~\bibnamefont {Imhof}}, \bibinfo {author} {\bibfnamefont
			{Q.}~\bibnamefont {Chen}}, \bibinfo {author} {\bibfnamefont {J.}~\bibnamefont
			{Raabe}}, \bibinfo {author} {\bibfnamefont {R.}~\bibnamefont {Betemps}},
		\bibinfo {author} {\bibfnamefont {D.}~\bibnamefont {Zimoch}}, \bibinfo
		{author} {\bibfnamefont {J.}~\bibnamefont {Krempasky}}, \bibinfo {author}
		{\bibfnamefont {X.}~\bibnamefont {Wang}}, \bibinfo {author} {\bibfnamefont
			{M.~P.~A.}\ \bibnamefont {Grioni}}, \ and\ \bibinfo {author} {\bibfnamefont
			{L.}~\bibnamefont {Patthey}},\ }\bibfield  {title} {\enquote {\bibinfo
			{title} {{{High-resolution soft X-ray beamline ADRESS at the Swiss Light
						Source for resonant inelastic X-ray scattering and angle-resolved
						photoelectron spectroscopies}}},}\ }\href {\doibase
		10.1107/S0909049510019862} {\bibfield  {journal} {\bibinfo  {journal} {J.
				Synchrotron Radiat.}\ }\textbf {\bibinfo {volume} {17}},\ \bibinfo {pages}
		{631--643} (\bibinfo {year} {2010})}\BibitemShut {NoStop}%
	\bibitem [{\citenamefont {M\aa{}nsson}\ \emph {et~al.}(2007)\citenamefont
		{M\aa{}nsson}, \citenamefont {Claesson}, \citenamefont {Karlsson},
		\citenamefont {Tjernberg}, \citenamefont {Pailh\'es}, \citenamefont {Chang},
		\citenamefont {Mesot}, \citenamefont {Shi}, \citenamefont {Patthey},
		\citenamefont {Momono} \emph {et~al.}}]{cleaver}%
	\BibitemOpen
	\bibfield  {author} {\bibinfo {author} {\bibfnamefont {M.}~\bibnamefont
			{M\aa{}nsson}}, \bibinfo {author} {\bibfnamefont {T.}~\bibnamefont
			{Claesson}}, \bibinfo {author} {\bibfnamefont {U.~O.}\ \bibnamefont
			{Karlsson}}, \bibinfo {author} {\bibfnamefont {O.}~\bibnamefont {Tjernberg}},
		\bibinfo {author} {\bibfnamefont {S.}~\bibnamefont {Pailh\'es}}, \bibinfo
		{author} {\bibfnamefont {J.}~\bibnamefont {Chang}}, \bibinfo {author}
		{\bibfnamefont {J.}~\bibnamefont {Mesot}}, \bibinfo {author} {\bibfnamefont
			{M.}~\bibnamefont {Shi}}, \bibinfo {author} {\bibfnamefont {L.}~\bibnamefont
			{Patthey}}, \bibinfo {author} {\bibfnamefont {N.}~\bibnamefont {Momono}},
		\emph {et~al.},\ }\bibfield  {title} {\enquote {\bibinfo {title} {On-board
				sample cleaver},}\ }\href {\doibase https://doi.org/10.1063/1.2756754}
	{\bibfield  {journal} {\bibinfo  {journal} {Rev. Sci. Instrum.}\ }\textbf
		{\bibinfo {volume} {78}},\ \bibinfo {pages} {076103} (\bibinfo {year}
		{2007})}\BibitemShut {NoStop}%
	\bibitem [{\citenamefont {Migliori}\ \emph {et~al.}(1990)\citenamefont
		{Migliori}, \citenamefont {Visscher}, \citenamefont {Brown}, \citenamefont
		{Fisk}, \citenamefont {Cheong}, \citenamefont {Alten}, \citenamefont
		{Ahrens}, \citenamefont {Kubat-Martin}, \citenamefont {Maynard},
		\citenamefont {Huang}, \citenamefont {Kirk}, \citenamefont {Gillis},
		\citenamefont {Kim},\ and\ \citenamefont {Chan}}]{MiglioriPRB1990}%
	\BibitemOpen
	\bibfield  {author} {\bibinfo {author} {\bibfnamefont {A.}~\bibnamefont
			{Migliori}}, \bibinfo {author} {\bibfnamefont {W.~M.}\ \bibnamefont
			{Visscher}}, \bibinfo {author} {\bibfnamefont {S.~E.}\ \bibnamefont {Brown}},
		\bibinfo {author} {\bibfnamefont {Z.}~\bibnamefont {Fisk}}, \bibinfo {author}
		{\bibfnamefont {S.-W.}\ \bibnamefont {Cheong}}, \bibinfo {author}
		{\bibfnamefont {B.}~\bibnamefont {Alten}}, \bibinfo {author} {\bibfnamefont
			{E.~T.}\ \bibnamefont {Ahrens}}, \bibinfo {author} {\bibfnamefont {K.~A.}\
			\bibnamefont {Kubat-Martin}}, \bibinfo {author} {\bibfnamefont {J.~D.}\
			\bibnamefont {Maynard}}, \bibinfo {author} {\bibfnamefont {Y.}~\bibnamefont
			{Huang}}, \bibinfo {author} {\bibfnamefont {D.~R.}\ \bibnamefont {Kirk}},
		\bibinfo {author} {\bibfnamefont {K.~A.}\ \bibnamefont {Gillis}}, \bibinfo
		{author} {\bibfnamefont {H.~K.}\ \bibnamefont {Kim}}, \ and\ \bibinfo
		{author} {\bibfnamefont {M.~H.~W.}\ \bibnamefont {Chan}},\ }\bibfield
	{title} {\enquote {\bibinfo {title} {Elastic constants and specific-heat
				measurements on single crystals {La}$_{2}${CuO}$_{4}$},}\ }\href {\doibase
		10.1103/physrevb.41.2098} {\bibfield  {journal} {\bibinfo  {journal} {Phys.
				Rev. B}\ }\textbf {\bibinfo {volume} {41}},\ \bibinfo {pages} {2098--2102}
		(\bibinfo {year} {1990})}\BibitemShut {NoStop}%
	\bibitem [{\citenamefont {Ledbetter}(1981)}]{LedbetterJAP1981}%
	\BibitemOpen
	\bibfield  {author} {\bibinfo {author} {\bibfnamefont {H.~M.}\ \bibnamefont
			{Ledbetter}},\ }\bibfield  {title} {\enquote {\bibinfo {title}
			{Stainless-steel elastic constants at low temperatures},}\ }\href {\doibase
		10.1063/1.329644} {\bibfield  {journal} {\bibinfo  {journal} {J. Appl.
				Phys.}\ }\textbf {\bibinfo {volume} {52}},\ \bibinfo {pages} {1587--1589}
		(\bibinfo {year} {1981})}\BibitemShut {NoStop}%
	\bibitem [{\citenamefont {Overton}\ and\ \citenamefont
		{Gaffney}(1955)}]{OvertonPR1998}%
	\BibitemOpen
	\bibfield  {author} {\bibinfo {author} {\bibfnamefont {W.~C.}\ \bibnamefont
			{Overton}}\ and\ \bibinfo {author} {\bibfnamefont {J.}~\bibnamefont
			{Gaffney}},\ }\bibfield  {title} {\enquote {\bibinfo {title} {Temperature
				variation of the elastic constants of cubic elements. {I}. copper},}\ }\href
	{\doibase 10.1103/PhysRev.98.969} {\bibfield  {journal} {\bibinfo  {journal}
			{Phys. Rev.}\ }\textbf {\bibinfo {volume} {98}},\ \bibinfo {pages} {969--977}
		(\bibinfo {year} {1955})}\BibitemShut {NoStop}%
\end{thebibliography}
\end{document}